\documentclass{fundam}
\issue{65(1-2), 61-86, 2005}

\usepackage{t1enc} 
\usepackage{amsmath}
\usepackage{latexsym}
\usepackage{theorem}
\usepackage{amssymb}
\usepackage{url}

\newcommand{\comment}[1]{}
\newcommand{\bu}{$\bullet$}

\newcommand{\moins}{\setminus}
\newcommand{\vide}{\emptyset}
\newcommand{\ie}{{\em i.e.} }
\newcommand{\eg}{{\em e.g.} }

\newcommand{\xu}{\{x\to u\}}

\newcommand{\xS}{\{x\to S\}}

\newcommand{\vyu}{\{\vy\to\vu\}}

\newcommand{\vxl}{\{\vx\to\vl\}}

\newcommand{\vxu}{\{\vx\to\vu\}}

\newcommand{\dom}{\mr{dom}}

\newcommand{\FV}{\mr{FV}}
\newcommand{\pos}{\mr{Pos}}

\renewcommand{\a}{\rightarrow}
\newcommand{\A}{\Rightarrow}
\renewcommand{\aa}{\leftrightarrow}

\newcommand{\ad}{\downarrow}

\renewcommand{\to}{\mapsto}

\newcommand{\ps}[1]{{\langle #1\rangle}}

\newcommand{\I}[1]{[\![#1]\!]}

\newcommand{\all}{\forall}
\newcommand{\ou}{\vee}

\newcommand{\et}{\wedge}
\newcommand{\non}{\neg}
\newcommand{\st}{\star}
\newcommand{\B}{\Box} 
\renewcommand{\th}{\vdash}

\newcommand{\sle}{\subseteq}

\newcommand{\tgt}{\rhd}

\renewcommand{\b}{\beta}
\newcommand{\g}{\gamma}
\newcommand{\G}{\Gamma}
\renewcommand{\d}{\delta}
\newcommand{\D}{\Delta}
\newcommand{\ep}{\epsilon}
\newcommand{\vep}{\varepsilon}

\renewcommand{\t}{\theta}

\newcommand{\io}{\iota}
\newcommand{\ka}{\kappa}
\newcommand{\la}{\lambda}

\renewcommand{\r}{\rho}
\newcommand{\s}{\sigma}

\newcommand{\Up}{\Upsilon}
\newcommand{\vphi}{\varphi}
\newcommand{\w}{\omega}

\newcommand{\mi}{\mathit}
\newcommand{\mc}{\mathcal}

\newcommand{\mr}{\mathrm}

\newcommand{\cA}{\mc{A}}
\newcommand{\cB}{\mc{B}}
\newcommand{\cC}{\mc{C}}
\newcommand{\cD}{\mc{D}}

\newcommand{\cF}{\mc{F}}
\newcommand{\cG}{\mc{G}}

\newcommand{\cK}{\mc{K}}

\newcommand{\cN}{\mc{N}}
\newcommand{\cO}{\mc{O}}
\newcommand{\cP}{\mc{P}}

\newcommand{\cR}{\mc{R}}
\newcommand{\cS}{\mc{S}}
\newcommand{\cT}{\mc{T}}

\newcommand{\cX}{\mc{X}}

\newcommand{\va}{{\vec{a}}}

\newcommand{\vc}{{\vec{c}}}

\newcommand{\vf}{{\vec{f}}}

\newcommand{\vl}{{\vec{l}}}
\newcommand{\vm}{{\vec{m}}}

\newcommand{\vq}{{\vec{q}}}

\newcommand{\vt}{{\vec{t}}}
\newcommand{\vu}{{\vec{u}}}
\newcommand{\vv}{{\vec{v}}}
\newcommand{\vw}{{\vec{w}}}
\newcommand{\vx}{{\vec{x}}}
\newcommand{\vy}{{\vec{y}}}
\newcommand{\vz}{{\vec{z}}}

\newcommand{\vA}{{\vec{A}}}
\newcommand{\vB}{{\vec{B}}}
\newcommand{\vC}{{\vec{C}}}
\newcommand{\vD}{{\vec{D}}}

\newcommand{\vQ}{{\vec{Q}}}
\newcommand{\vR}{{\vec{R}}}
\newcommand{\vS}{{\vec{S}}}
\newcommand{\vT}{{\vec{T}}}
\newcommand{\vU}{{\vec{U}}}
\newcommand{\vV}{{\vec{V}}}
\newcommand{\vW}{{\vec{W}}}

\newenvironment{rul}
  {$\begin{array}{rcl}}
  {\end{array}$}

\newenvironment{rew}[1][~~\a~~]
  {$\begin{array}{r@{#1}l}}
  {\end{array}$}
\newenvironment{rewc}[1][~~\a~~]
  {\begin{center}\begin{rew}[#1]}
  {\end{rew}\end{center}}

  {\end{array}$\end{center}}

\newenvironment{typc}[1][\,:\,]{\begin{rewc}[#1]}{\end{rewc}}

{\theorembodyfont{\rmfamily} 

}

\newenvironment{lstgeneric}[2]
  {\begin{list}{#1}{\topsep=.5mm\itemsep=.5mm\parsep=0mm%
    \itemindent=-3ex\labelsep=1ex\labelwidth=0ex #2}}
  {\end{list}}

\newenvironment{lst}[1]
  {\begin{lstgeneric}{#1}{\itemindent=-1ex}}
  {\end{lstgeneric}}

\newenvironment{enumi}[1]
  {\begin{lstgeneric}{}{\usecounter{enumi}\leftmargin=7mm%
    }}
  {\end{lstgeneric}}

\newenvironment{bfenumii}[1]
  {\begin{lstgeneric}{}{\usecounter{enumii}\leftmargin=7mm%
    }}
  {\end{lstgeneric}}

\newcommand{\tf}{{\tau_f}}
\newcommand{\tF}{{\tau_F}}
\newcommand{\tg}{{\tau_g}}

\newcommand{\tC}{{\tau_C}}

\newcommand{\FVB}{\FV^\B}

\newcommand{\XB}{\cX^\B}
\newcommand{\Xs}{\cX^\st}

\newcommand{\SN}{\cS\cN}

\newcommand{\NF}{\cN\cF}

\newcommand{\mon}{\mr{Mon}}
\newcommand{\rec}{\cR ec}
\newcommand{\acc}{\mr{Acc}}
\newcommand{\cons}{{\cC ons}}

\newcommand{\br}{{\b\cR}}

\newcommand{\fin}{{\mi{fin}}}
\newcommand{\femp}{\mi{empty}}
\newcommand{\fadd}{\mi{add}}

\newcommand{\JMeq}{\mi{JMeq}}
\newcommand{\refl}{\mi{refl}}

\newcommand{\ai}{\a_\io}
\newcommand{\abe}{\a_{\b\eta}}
\newcommand{\abi}{\a_{\b\io}}
\newcommand{\abr}{\a_{\b\cR}}
\newcommand{\ab}{\a_\b}
\newcommand{\ar}{\a_\cR}
\renewcommand{\ae}{\a_\eta}

\newcommand{\XI}{\{X\!\to\!I\}}
\newcommand{\aip}{\a_{\io'}}
\newcommand{\abip}{\a_{\b\io'}}
\newcommand{\thu}{\,{\th\!\!\!\!_{_\Up}}\,}

\newcommand{\FB}{\cF^\B}
\newcommand{\CFB}{\cC\FB}
\newcommand{\CFBI}{\CFB_{intro}}
\newcommand{\CFBE}{\CFB_{elim}}

\newcommand{\we}{{W\!Elim_I}}
\newcommand{\se}{{S\!Elim_I^Q}}
\newcommand{\vxa}{\{\vx\to\va\}}

\begin{document}


\title{Inductive types in the Calculus of Algebraic Constructions}

\author{Fr\'ed\'eric Blanqui\\
LORIA \& INRIA\\
615 rue du Jardin Botanique, BP 101, 54602 Villers-l\`es-Nancy, France\\
\em\url{http://www.loria.fr/~blanqui/} - \url{blanqui@loria.fr}}

\maketitle

\setcounter{page}{1}
\issue{?}
\runninghead{F. Blanqui}{Inductive types in the Calculus of Algebraic Constructions}

\begin{abstract}
In a previous work, we proved that an important part of the Calculus
of Inductive Constructions (CIC), the basis of the Coq proof
assistant, can be seen as a Calculus of Algebraic Constructions (CAC),
an extension of the Calculus of Constructions with functions and
predicates defined by higher-order rewrite rules. In this paper, we
prove that almost all CIC can be seen as a CAC, and that it can be
further extended with non-strictly positive types and
inductive-recursive types together with non-free constructors and
pattern-matching on defined symbols.
\end{abstract}


\section{Introduction}
\label{sec-intro}

There has been different proposals for defining inductive
types\footnote{All over the paper, by ``inductive types'', we also
mean inductively defined predicates or families of types.} and
functions in typed systems. In Girard's polymorphic $\la$-calculus or
in the Calculus of Constructions (CC) \cite{coquand88ic}, data types
and functions can be formalized by using impredicative encodings,
difficult to use in practice, and computations are done by
$\b$-reduction only. In Martin-L\"of's type theory or in the Calculus
of Inductive Constructions (CIC) \cite{coquand88colog}, inductive
types and their induction principles are first-class objects,
functions can be defined by induction and computations are done by
$\io$-reduction, the rules for cut-elimination in inductive
proofs. For instance, for the type $nat$ of natural numbers, the
recursor\footnote{$(x:T)P$ is a usual type-theoretic notation for the
dependent product or universal quantification ``for all $x$ of type
$T$, $P$''.} $rec:(P:nat\A\st)(u:P0)(v:(n:nat)$ $Pn\A P(sn))(n:nat)Pn$
is defined by the following $\io$-rules:

\begin{rewc}[~~\a_\io~~]
rec~P~u~v~0 & u\\
rec~P~u~v~(s~n) & v~n~(rec~P~u~v~n)\\
\end{rewc}

Finally, in the algebraic setting \cite{dershowitz90book}, functions
are defined by using rewrite rules and computations are done by
applying these rules. Since both $\b$-reduction and $\io$-reduction
are particular cases of higher-order rewriting \cite{klop93tcs},
proposals soon appeared for integrating all these approaches. Starting
with \cite{jouannaud91lics,barbanera94lics}, this objective culminated
with \cite{blanqui01lics,blanqui01thesis,blanqui05mscs} in which an
important part of CIC (described in \cite{blanqui01thesis}) can be
seen as a Calculus of Algebraic Constructions (CAC), an extension of
CC with functions and predicates defined by higher-order rewrite
rules. In this paper, we go one step further in this direction,
capture almost all CIC and extend it with non-strictly positive
inductive types and inductive recursive types \cite{dybjer00jsl}.


Let us see two examples of recursors that are allowed in CIC but not
in CAC \cite{paulin01pc}. The first example is a third-order
definition of finite sets of natural numbers (represented as
predicates over $nat$):

\begin{typc}[\,]
\fin: & (nat\A\st)\A\st\\
\femp: & \fin([y:nat]\bot)\\
\fadd: & (x:nat)(p:nat\A\st)\fin\,p\A \fin([y:nat]y=x\ou(p~y))\\
rec: & (Q:(nat\A\st)\A\st)Q([y:nat]\bot)\\
& \A ((x:nat)(p:nat\A\st)\fin\,p\A Qp\A Q([y:nat]y=x\ou(p~y)))\\
& \A (p:nat\A\st)\fin\,p\A Qp\\
\end{typc}

\noindent
where $\bot$ is the false proposition and the {\em weak} recursor
$rec$, \ie the recursor for defining objects, is defined by the rules:

\begin{rewc}
rec~Q~u~v~p'~\femp & u\\
rec~Q~u~v~p'~(\fadd~x~p~h) & v~x~p~h~(rec~Q~u~v~p~h)\\
\end{rewc}

The problem comes from the fact that, in the output type of $\fadd$,
$\fin([y:nat]y=x\ou(p~y))$, the predicate $p$ is not parameter of
$\fin$. This is why the corresponding {\em strong} recursor, \ie the
recursor for defining types or predicates, is not allowed in CIC ($p$
could be ``bigger'' than $\fin$) \cite{coquand86lics}. This can be
generalized to any big/impredicative dependent type, that is, to any
type having a constructor with a predicate argument which is not a
parameter. Formally, this condition, called {\bf(I6)} in
\cite{blanqui05mscs}, {\em safeness} in \cite{stefanova98thesis} and
{\em $\st$-dependency for constructors} in \cite{walukiewicz03jfp},
can be stated as follows:

\begin{definition}[I6]
\label{def-i6}
If $C:(\vz:\vV)\st$ is a type and $c:(\vx:\vT)C\vv$ is a constructor
of $C$ then, for all predicate variable $x$ occurring in some $T_j$,
there is some argument $v_{\io_x}=x$.
\end{definition}


The second example is John Major's equality which is intended to equal
terms of different types \cite{mcbride99thesis}:

\begin{typc}[\,]
\JMeq: & (A:\st)A\A(B:\st)B\A\st\\
\refl: & (C:\st)(x:C)(\JMeq~C~x~C~x)\\
rec: & (A:\st)(x:A)(P:(B:\st)B\A\st)(P~A~x)\\
& \A(B:\st)(y:B)(\JMeq~A~x~B~y)\A (P~B~y)\\
\end{typc}

\noindent
where $rec$ is defined by the rule:

\begin{rewc}
rec~C~x~P~h~C~x~(\refl~C~x) & h\\
\end{rewc}

\noindent
Here, the problem comes from the fact that, in the output type of
$\refl$, the argument for $B$ is equal to the argument for $A$. This
can be generalized to any polymorphic type having a constructor with
two equal type parameters. From a rewriting point of view, this is
like having pattern-matching or non-linearities on predicate
arguments, which is known to create inconsistencies in some cases
\cite{harper99ipl}. A similar restriction called {\em $\st$-dependency
for function symbols} also appears in \cite{walukiewicz03jfp}.

\begin{definition}[Safeness]
\label{def-safe}
A rule $f\vl\a r$ with $f:(\vx:\vT)U$ is {\em safe} if:

\begin{lst}{--}
\item for all predicate argument $x_i$, $l_i$ is a variable,
\item if $x_i$ and $x_j$ are two distinct predicate arguments,
then $l_i\neq l_j$.
\end{lst}

\noindent
An inductive type is {\em safe} if the corresponding $\io$-rules are
safe.
\end{definition}


By using what is called in Matthes' terminology \cite{matthes98thesis}
an {\em elimination-based} interpretation instead of the {\em
introduction-based} interpretation that we used in
\cite{blanqui05mscs}, we prove that weak recursors for types like $\fin$
or $\JMeq$ can be accepted, hence that CAC subsumes CIC almost
completely. The only condition we could not get rid of is the safeness
condition for predicate-level rewrite rules. So, we do not accept
strong elimination on $\JMeq$ (strong elimination for $\fin$ is
allowed neither in CIC nor in CAC \cite{coquand86lics}). On the other
hand, we prove that CAC and CIC can be easily extended to non-strictly
positive types (Section \ref{sec-pos}) and to inductive-recursive
types (Section \ref{sec-indrec}) \cite{dybjer00jsl}.


\section{The Calculus of Inductive Constructions (CIC)}
\label{sec-cic-syntax}

We assume the reader familiar with typed $\la$-calculi
\cite{barendregt92book}. In this section, we present CIC as defined in
\cite{werner94thesis}.  In order to type the strong elimination schema
in a polymorphic way, which is not possible in CC, Werner uses a
slightly more general Pure Type System (PTS)
\cite{barendregt92book}. CC is the PTS with the sorts
$\cS=\{\st,\B\}$, the axioms $\cA=\{(\st,\B)\}$ and the rules
$\cB=\{(s_1,s_2,s_3)\in\cS^3~|~s_2=s_3\}$. Werner extends it by adding
the sort $\triangle$, the axiom $(\B,\triangle)$ and the rules
$(\st,\triangle,\triangle)$ and $(\B,\triangle,\triangle)$. In fact,
he denotes $\st$ by {\em Set}, $\B$ by {\em Type} and $\triangle$ by
{\em Extern}. The sort $\st$ denotes the universe of types and
propositions, and the sort $\B$ denotes the universe of predicate
types (also called {\em kinds}). For instance, the type $nat$ of
natural numbers is of type $\st$, $\st$ itself is of type $\B$ and
$nat\A\st$, the type of predicates over $nat$, is of type $\B$. Then,
Werner adds terms for representing inductive types, their constructors
and the definitions by recursion on these types:

\begin{lst}{\bu}
\item {\bf Inductive types.} An inductive type is denoted by
$I=Ind(X:A)\{\vC\}$ where $\vC$ is an ordered sequence of terms for
the types of the constructors of $I$. For instance,
$Nat=Ind(X:\st)\{X,X\A X\}$ represents the type of natural numbers (in
fact, any type isomorphic to the type of natural numbers). The term
$A$ must be of the form $(\vx:\vA)\st$ and the $C_i$'s of the form
$(\vz:\vB)X\vm$ with no $X$ in $\vm$. Furthermore, the inductive types
must be strictly positive. In CIC, this means that, if
$C_i=(\vz:\vB)X\vm$ then, for all $j$, either $X$ does not occur in
$B_j$, or $B_j$ is of the form $(\vy:\vD)X\vq$ and $X$ occurs neither
in $\vD$ nor in $\vq$.
  
\item {\bf Constructors.} The $i$-th constructor
of an inductive type $I$ is denoted by $Constr(i,I)$. For instance,
$Constr(1,Nat)$ represents zero and $Constr(2,Nat)$ represents the
successor function.

\item {\bf Definitions by recursion.} A definition by recursion
on an inductive type $I$ is denoted by $Elim(I,Q,\va$, $c)$ where $Q$
is the type of the result, $\va$ the arguments of $I$ and $c$ a term
of type $I\va$. The strong elimination (\ie when $Q$ is a predicate
type) is restricted to {\em small} inductive types, that is, to the
types whose constructors have no other predicate arguments than the
ones that their type have. Formally, an inductive type
$I=Ind(X:A)\{\vC\}$ is {\em small} if all the types of its
constructors are small, and a constructor type $C=(\vz:\vB)X\vm$ is
{\em small} if $\vz$ are object variables (this means that the
predicate arguments must be part of the environment in which they are
typed; they cannot be part of $\vC$).
\end{lst}


For defining the reduction relation associated with $Elim$, called
{\em $\io$-reduction} and denoted by $\ai$, and the typing rules of
these inductive constructions (see Figure~\ref{fig-th-cic}), it is
necessary to introduce a few definitions. Let $C$ be a constructor
type. We define $\D\{I,X,C,Q,c\}$ as follows:

\begin{lst}{--}
\item $\D\{I,X,X\vm,Q,c\}= Q\vm c$
\item $\D\{I,X,(z:B)D,Q,c\}= (z:B)\D\{I,X,D,Q,cz\}$
if $X$ does not occur in $B$
\item $\D\{I,X,(z:B)D,Q,c\}=
(z:B\XI)((\vy:\vD)Q\vq\,(z\vy))\A\D\{I,X,D,Q,cz\}$\\
if $B=(\vy:\vD)X\vq$
\end{lst}

Then, the {\em $\io$-reduction} is defined by the rule:

\begin{center}
$Elim(I,Q,\vx,Constr(i,I')\vz)\{\vf\} ~\ai~
\D[I,X,C_i,f_i,FunElim(I,Q,\vf)]\vz$
\end{center}

\noindent
where $I=Ind(X:A)\{\vC\}$, $FunElim(I,Q,\vf)=
[\vx:\vA][y:I\vx]Elim(I,Q,\vx,y)\{\vf\}$ and $\D[I,X,C$, $f,F]$ is
defined as follows:

\begin{lst}{--}
\item $\D[I,X,X\vm,f,F]= f$
\item $\D[I,X,(z:B)D,f,F]= [z:B]\D[I,X,D,fz,F]$
if $X$ does not occur in $B$
\item $\D[I,X,(z:B)D,f,F]= [z:B\XI]\D[I,X,D,fz[\vy:\vD](F\vq\,(z\vy)),F]$
if $B=(\vy:\vD)X\vq$
\end{lst}

Finally, in the type conversion rule (Conv), in addition to
$\b$-reduction and $\io$-reduction, Werner considers $\eta$-reduction:
$[x:T]ux\ae u$ if $x$ does not occur in $u$. The relation
$\aa^*_{\b\eta\io}$ is the reflexive, symmetric and transitive closure
of $\a_{\b\eta\io}$. Note that, since $\abe$ is not confluent on badly
typed terms \cite{nederpelt73thesis}, considering $\eta$-reduction
creates important difficulties.


\begin{figure}[ht]
\begin{center}
\caption{Typing rules for inductive constructions in CIC\label{fig-th-cic}}

\begin{tabular}{r@{~}c}
\\(Ind) & $\cfrac{
\begin{array}{c}
A=(\vx:\vA)\st \quad \G\th A:\B \quad \all i,\, \G,X:A\th C_i:\st\\
I=Ind(X:A)\{\vC\} \mbox{ is strictly positive}\\
\end{array}}
{\G\th I:A}$\\

\\(Constr) & $\cfrac{I=Ind(X:A)\{\vC\} \quad \G\th I:T}
{\G\th Constr(i,I):C_i\XI}$\\

\\($\st$-Elim) & $\cfrac{
\begin{array}{c}
A=(\vx:\vA)\st \quad I=Ind(X:A)\{\vC\} \quad \G\th Q:(\vx:\vA)I\vx\A\st\\
T_i=\D\{I,X,C_i,Q,Constr(i,I)\}\\
\all j,\, \G\th a_j:A_j\vxa \quad \G\th c:I\va
\quad \all i,\, \G\th f_i:T_i\\
\end{array}}
{\G\th Elim(I,Q,\va,c)\{\vf\}:Q\va c}$\\

\\($\B$-Elim) & $\cfrac{
\begin{array}{c}
A=(\vx:\vA)\st \quad I=Ind(X:A)\{\vC\} \mbox{ is small}\quad
\G\th Q:(\vx:\vA)I\vx\A\B\\
T_i= \D\{I,X,C_i,Q,Constr(i,I)\}\\
\all j,\,\G\th a_j:A_j\vxa \quad \G\th c:I\va
\quad \all i,\, \G\th f_i:T_i\\
\end{array}}
{\G\th Elim(I,Q,\va,c)\{\vf\}:Q\va c}$\\

\\(Conv) & $\cfrac{\G\th t:T\quad T \aa_{\b\eta\io}^* T'\quad \G\th T':s}
{\G\th t:T'}$\\
\end{tabular}
\end{center}
\end{figure}


\section{The Calculus of Algebraic Constructions (CAC)}
\label{sec-cac}

We assume the reader familiar with rewriting
\cite{dershowitz90book}. The Calculus of Algebraic Constructions (CAC)
\cite{blanqui05mscs} simply extends CC with a set $\cF$ of {\em
symbols} and a set $\cR$ of {\em rewrite rules} (see Definition
\ref{def-rule}).

\begin{definition}[Terms]
\label{def-terms}
The set $\cT$ of CAC terms is inductively defined as follows:

\begin{center}
$t,u\in\cT ::= s ~|~ x ~|~ f ~|~ [x:t]u ~|~ tu ~|~ (x:t)u$
\end{center}

\noindent
where $s\in\cS=\{\st,\B\}$ is a {\em sort}, $x\in\cX$ is a {\em
variable}, $f\in\cF$ is a {\em symbol}, $[x:t]u$ is an {\em
abstraction}, $tu$ is an {\em application}, and $(x:t)u$ is a {\em
dependent product}, written $t\A u$ if $x$ does not freely occur in
$u$. As usual, terms are considered up to $\alpha$-conversion, \ie up
to sort-preserving renaming of bound variables. A term $t$ is {\em of
the form} a term $u$ if $t$ is $\alpha$-convertible to $u\s$ for some
substitution $\s$.
\end{definition}

We denote by $\FV(t)$ the set of variables that freely occur in $t$,
by $\pos(t)$ the set of Dewey's positions in $t$ (words on strictly
positive integers), by $t|_p$ the subterm of $t$ at position $p$, by
$\pos(x,t)$ the set of positions $p\in\pos(t)$ such that $t|_p$ is a
free occurrence of $x$ in $t$, and by $\dom(\t)=
\{x\in\cX~|~ x\t\neq x\}$ the {\em domain} of a substitution $\t$.
Let $\vt$ denote a sequence of terms $t_1\ldots t_n$ of length
$|\vt|=n\ge 0$.

Every $x\in\cX\cup\cF$ is equipped with a sort $s_x$. We denote by
$\cX^s$ (resp. $\cF^s$) the set of variables (resp. symbols) of sort
$s$. Let $\FV^s(t)=\FV(t)\cap\cX^s$ and
$\dom^s(\t)=\dom(\t)\cap\cX^s$. A variable or a symbol of sort $\st$
(resp. $\B$) is an {\em object} (resp. a {\em predicate}).

Although terms and types are mixed in Definition \ref{def-terms}, we
can distinguish the following three disjoint sub-classes where
$t\in\cT$ denotes any term:

\begin{lst}{--}
\item objects: $o\in\cO ::= x\in\cX^\st ~|~ f\in\cF^\st ~|~ [x:t]o ~|~ ot$
\item predicates: $P\in\cP ::= x\in\cX^\B ~|~ f\in\cF^\B ~|~ [x:t]P ~|~ Pt
~|~ (x:t)P$
\item predicate types or kinds: $K\in\cK ::= \st ~|~ (x:t)K$
\end{lst}


\begin{figure}[ht]
\centering
\caption{Typing rules of CAC\label{fig-typ}}
\begin{tabular}{rcc}
\\ (ax) & $\th\st:\B$\\

\\ (symb) & $\cfrac{\th\tf:s_f}{\th f:\tf}$\\

\\ (var) & $\cfrac{\G\th T:s_x}{\G,x:T\th x:T}$
& $(x\notin\dom(\G))$\\

\\ (weak) & $\cfrac{\G\th t:T \quad \G\th U:s_x}{\G,x:U\th t:T}$
& $(x\notin\dom(\G))$\\

\\ (prod) & $\cfrac{\G\th U:s \quad \G,x:U \th V:s'}
{\G\th (x:U)V:s'}$\\

\\ (abs) & $\cfrac{\G,x:U \th v:V \quad \G\th (x:U)V:s}
{\G\th [x:U]v:(x:U)V}$\\

\\ (app) & $\cfrac{\G\th t:(x:U)V \quad \G\th u:U}
{\G\th tu:V\xu}$\\

\\ (conv) & $\cfrac{\G\th t:T \quad \G\th T':s}{\G\th t:T'}$ & ($T\ad_\br T'$)
\\
\end{tabular}
\end{figure}


\begin{definition}[Precedence]
We assume given a total quasi-ordering $\ge$ on symbols whose strict
part $>=\ge\moins\le$ is well-founded, and let ${\simeq}={\ge\cap\le}$
be its associated equivalence relation. A symbol $f$ is {\em smaller}
(resp. {\em strictly smaller}) than a symbol $g$ iff $f\le g$
(resp. $f<g$). A symbol $f$ is {\em equivalent} to a symbol $g$ iff
$f\simeq g$.
\end{definition}


\begin{definition}[Rewrite rule]
\label{def-rule}
The terms only built from variables and applications of the form
$f\vt$ are called {\em algebraic}. A {\em rewrite rule} is a pair $l\a
r$ such that:

\begin{lst}{--}
\item $l$ is algebraic,
\item $l$ is not a variable,
\item $\FV(r)\sle\FV(l)$,
\item every symbol occurring in $r$ is smaller than $f$.
\end{lst}

\noindent
The rewrite relation $\ar$ induced by $\cR$ is the smallest relation
containing $\cR$ and stable by context and substitution: $t\ar t'$ iff
there exist $p\in\pos(t)$, $l\a r\in\cR$ and $\s$ such that
$t=t[l\s]_p$ and $t'=t[r\s]_p$. A symbol $f$ with no rule $f\vl\a
r\in\cR$ is {\em constant}, otherwise it is (partially) {\em
defined}. Let $\cC\cF^s$ (resp. $\cD\cF^s$) be the set of constant
(resp. defined) symbols of sort $s$.
\end{definition}


\begin{definition}[Typing]
\label{def-typing}
Every $f\in\cF$ is equipped with a {\em type} $\tf$ such that:

\begin{lst}{--}
\item $\tf$ is a closed term of the form $(\vx:\vT)U$ with $U$ distinct
from a product,
\item every symbol occurring in $\tf$ is strictly smaller than $f$,
\item for every rule $f\vl\a r\in\cR$, we have $|\vl|\le|\vx|$.
\end{lst}

\noindent
A {\em constructor} is any symbol $f$ whose type is of the form
$(\vy:\vU)C\vv$ with $C\in\CFB$. Let $\cons$ be the set of
constructors. A typing {\em environment} is a sequence of
variable-type pairs. Given $f$ of type $(\vx:\vT)U$, we denote by
$\G_f$ the environment $\vx:\vT$.

The typing relation of CAC is the relation $\th$ defined in Figure
\ref{fig-typ}. Let $\th_g$ (resp. $\th_g^<$) be the typing relation
defined by the rules of Figure \ref{fig-typ} with the side condition
$f\le g$ (resp. $f<g$) in the (symb) rule.
\end{definition}


In comparison with CC, we added the rule (symb) for typing symbols
and, in the rule (conv), we replaced $\ad_\b$ by $\ad_\br$, where
$u\ad_\br v$ iff there exists a term $w$ such that $u\abr^* w$ and
$v\abr^* w$, $\abr^*$ being the reflexive and transitive closure of
$\a_\br=\ab\cup\ar$. This means that types having a common reduct are
identified and share the same proofs: any term of type $T$ is also of
type $T'$ if $T$ and $T'$ have a common reduct. For instance, a proof
of $P(2+2)$ is also a proof of $P(4)$ if $\cR$ contains the rules:

\begin{rewc}
x+0 & x\\
x+(s~y) & s~(x+y)\\
\end{rewc}

This decreases the size of proofs by an important factor, and
increases the automation as well. {\bf All over the paper, we assume
that $\a=\abr$ is confluent}. This is the case if, for instance, $\cR$
is left-linear and confluent \cite{muller92ipl}, like $\io$-reduction
is.

A substitution $\t$ {\em preserves typing from $\G$ to $\D$}, written
$\t:\G\leadsto\D$, if, for all $x\in\dom(\G)$, $\D\th x\t:x\G\t$,
where $x\G$ is the type associated to $x$ in $\G$. Type-preserving
substitutions enjoy the following important property: if $\G\th t:T$
and $\t:\G\leadsto\D$ then $\D\th t\t:T\t$ (Lemma 24 in
\cite{blanqui01thesis}).


For ensuring the {\em subject reduction} property (preservation of
typing under reduction, see Theorems 5 and 16 in
\cite{blanqui05mscs}), rules must satisfy the following conditions
(see Definition 3 in \cite{blanqui05mscs}):

\begin{definition}[Well-typed rules]
Every rule $f\vl\a r$ is assumed to be equipped with an environment
$\G$ and a substitution $\r$ such that, if $\tf=(\vx:\vT)U$ and
$\g=\vxl$, the following conditions are satisfied:

\begin{lst}{--}
\item $\G\th r:U\g\r$,
\item $\all\D,\s,T$, if $\D\th l\s:T$ then $\s:\G\leadsto\D$ and $\s\ad\r\s$.
\end{lst}
\end{definition}

The first condition is decidable under the quite natural restriction
that the typing of $r$ does not need the use of $f\vl\a r$. The other
conditions generally follow from the inversion of the judgment $\D\th
l\s:T$, and confluence for the condition $\s\ad\r\s$. Lemma 7 in
\cite{blanqui05mscs} gives sufficient conditions for deciding
that $\s:\G\leadsto\D$.

The substitution $\r$ allows to eliminate non-linearities only due to
typing. This makes rewriting more efficient and the proof of
confluence easier. For instance, the concatenation on polymorphic
lists (type $list:\st\A\st$ with constructors $nil:(A:\st)listA$ and
$cons:{(A:\st)}A\A listA\A listA$) of type $(A:\st)listA\A listA\A
listA$ can be defined by:

\begin{rewc}
app~A~(nil~A')~l' & l'\\
app~A~(cons~A'~x~l)~l' & cons~A~x~(app~A~x~l~l')\\
app~A~(app~A'~l~l')~l'' & app~A~l~(app~A~l'~l'')\\
\end{rewc}

\noindent
with $\G=A:\st,x:A,l:listA,l':listA$ and $\r=\{A'\to A\}$. Note that
the third rule has no counterpart in CIC. Although $app~A~(nil~A')$ is
not typable in $\G$ (since $A'\notin\dom(\G)$), it becomes typable if
we apply $\r$. This does not matter since, if an instance
$app~A\s~(nil~A'\s)$ is typable then, after the typing rules, $A\s$ is
convertible to $A'\s$. See \cite{blanqui05mscs} for details.


We now introduce some restrictions on predicate-level rewrite rules,
that generalize usual restrictions of strong elimination. Indeed, it
is well known that strong elimination on big inductive types may lead
to inconsistencies \cite{coquand86lics}.

\begin{definition}[Conditions on predicate-level rules]
\label{def-cond-pred}
\begin{lst}{--}
\item For all $F\in\cF^\B$, $F\vl\a r\in\cR$ and $x\in\FVB(r)$,
there is $\ka_x$ such that $l_{\ka_x}=x$.
\item Predicate-level rules have critical pairs with no rule.
\end{lst}
\end{definition}

The first condition means that one cannot do matching on predicate
arguments, hence that predicate variables are like parameters.

The condition on critical pairs, which is satisfied by CIC recursors,
allows us to define an interpretation for defined predicate symbols
easily (see Definition \ref{def-int-def}). However, we think that this
condition could be weakened. For instance, consider
$F:nat\A\st\A\st\A\st$ and the rules:

\begin{rewc}
F~0~A~B & B\\
F~(s~n)~A~B & A\A(F~n~A~B)\\
\end{rewc}

$(F~n~A~B)$ is the type of functions with $n$ arguments of type $A$
and output in $B$. So, it seems reasonable to allow rules derived from
inductive consequences of these first two rules, like for instance:

\begin{rewc}
F~(x+y)~A~B & F~x~A~(F~y~A~B)\\
\end{rewc}


We now prove a simple lemma saying that, for proving a property $P$
for every typing judgment $\G\th t:T$, one may proceed by well-founded
induction on the symbol precedence and prove that $P$ holds for every
typing judgment $\G\th_g t:T$ when it holds for every typing
judgment $\G\th_f t:T$ such that $f<g$.

\begin{lemma}
\label{lem-ind-prec}
We have (1) $\G\th t:T$ and every symbol occurring in $\G,t,T$ smaller
(resp. strictly smaller) than $g$ if and only if (2) $\G\th_g t:T$
(resp. $\G\th_g^< t:T$).
\end{lemma}

\begin{proof}
(1) $\A$ (2). One can easily prove by induction on $\G\th t:T$ that,
(*) if $\G\th t:T$ and every symbol occurring in $\G$ and $t$ is smaller
than $g$, then there exists $T'$ such that $T\a^* T'$ and $\G\th_g
t:T'$ (see Lemma 54 in \cite{blanqui01thesis}). In the (symb) case, it
uses the assumption that every symbol occurring in $\tf$ is strictly
smaller than $f$ (Definition \ref{def-typing}). In the (conv) case, it
uses confluence and the assumption that, for every rule $f\vl\a r$,
the symbols occurring in $r$ are smaller than $f$ (Definition
\ref{def-rule}). So, assume that $\G\th t:T$ and every symbol
occurring in $\G,t,T$ is smaller than $g$. By (*), there exists $T'$
such that $T\a^* T'$ and $\G\th_g t:T'$. By type correctness (Lemma 28
in \cite{blanqui01thesis}), either $T=\B$ or $\G\th T:s$. If $T=\B$
then $T'=T=\B$ and $\G\th_g t:T$. Now, if $\G\th T:s$ then, by (*)
again, $\G\th_g T:s$. Thus, by (conv), $\G\th_g t:T$. The same holds
with $\th_g^<$.

\noindent
(2) $\A$ (1). Easy induction on $\G\th_g t:T$.
\end{proof}

\begin{corollary}
\label{cor-typ}
If $\th g:\tg$ then $\th_g^<\tg:s_g$.
\end{corollary}

\begin{proof}
It follows from Lemma \ref{lem-ind-prec} and the assumption that, for
all $f$, every symbol occurring in $\tf$ is strictly smaller than $f$
(see Definition \ref{def-typing}).
\end{proof}


\section{Strong normalization}
\label{sec-sn}

Typed $\la$-calculi are generally proved strongly normalizing by using
Tait and Girard's technique of {\em reducibility candidates}
\cite{girard88book}. The idea of Tait, later extended by Girard to the
polymorphic $\la$-calculus, is to strengthen the induction
hypothesis. Instead of proving that every term is strongly
normalizable (set $\SN$), one associates to every type $T$ a set
$\I{T}\sle\SN$, the {\em interpretation} of $T$, and proves that every
term $t$ of type $T$ is {\em computable}, \ie belongs to
$\I{T}$. Hereafter, we follow the proof given in \cite{blanqui05mscs}
which greatly simplifies the one given in \cite{blanqui01thesis}. All
the definitions and properties of this section are taken from
\cite{blanqui05mscs}.


\begin{definition}[Reducibility candidates]
\label{def-cand}
We assume given a set $\cN\sle\cT$ of {\em neutral terms} satisfying
the following property: if $t\in\cN$ and $u\in\cT$ then $tu$ is not
head-reducible. We inductively define the complete lattice $\cR_t$ of
the interpretations for the terms of type $t$, the ordering $\le_t$ on
$\cR_t$, and the greatest element $\top_t\in\cR_t$ as follows.

\begin{lst}{--}
\item $\cR_t=\{\vide\}$, $\le_t=\sle$ and $\top_t=\vide$ if $t\neq\B$
and $t$ is not of the form $(\vx:\vT)\st$.

\item $\cR_s$ is the set of all subsets $R\sle\cT$ such that:
\begin{bfenumii}{R}
\item $R\sle\SN$ (strong normalization).
\item If $t\in R$ then $\a\!\!(t)=\{t'\in\cT~|~t\a t'\}\sle R$
(stability by reduction).
\item If $t\in\cN$ and $\a\!\!(t)\sle R$ then $t\in R$ (neutral terms).
\end{bfenumii}
Furthermore, $\le_s=\sle$ and $\top_s=\SN$.

\item $\cR_{(x:U)K}$ is the set of functions $R$ from $\cT\times\cR_U$
to $\cR_K$ such that $R(u,S)=R(u',S)$ whenever $u\a u'$,
$R\le_{(x:U)K} R'$ iff, for all $(u,S)\in\cT\times\cR_U$, $R(u,S)\le_K
R'(u,S)$, and $\top_{(x:U)K}(u,S)=\top_K$.
\end{lst}
\end{definition}

The exact definition of $\cN$ is not necessary at this
stage. Moreover, the choice of $\cN$ may depend on the way predicate
symbols are interpreted. The set that we will choose is given in
Definition \ref{def-neutr}.

Note that $\cR_t=\cR_{t'}$ whenever $t\a t'$ (Lemma 34 in
\cite{blanqui05mscs}). The proof that $(\cR_t,\le_t)$ is a complete
lattice is given in Lemma 35 in \cite{blanqui05mscs}.


\begin{definition}[Interpretation schema]
\label{def-schema-int}
A {\em candidate assignment} is a function $\xi$ from $\cX$ to
$\bigcup \,\{\cR_t ~|~ t\in\cT\}$. An assignment $\xi$ {\em validates}
an environment $\G$, $\xi\models\G$, if, for all $x\in\dom(\G)$,
$x\xi\in \cR_{x\G}$. An {\em interpretation} for a symbol $f$ is an
element of $\cR_\tf$. An {\em interpretation} for a set $\cG$ of
symbols is a function which, to every symbol $g\in\cG$, associates an
interpretation for $g$. The {\em interpretation} of a term $t$
w.r.t. a candidate assignment $\xi$, an interpretation $I$ for $\cF$
and a substitution $\t$, is defined by induction on $t$ as follows:

\begin{lst}{\bu}
\item $\I{t}^I_{\xi,\t}= \top_t$ if $t$ is an object or a sort,
\item $\I{x}^I_{\xi,\t}= x\xi$,
\item $\I{f}^I_{\xi,\t}= I_f$,
\item $\I{(x:U)V}^I_{\xi,\t}= \{t\in\cT~|~ \all u\in\I{U}^I_{\xi,\t},
\all S\in\cR_U, tu\in\I{V}^I_{\xi_x^S,\t_x^u}\}$,
\item $\I{[x:U]v}^I_{\xi,\t}(u,S)= \I{v}^I_{\xi_x^S,\t_x^u}$,
\item $\I{tu}^I_{\xi,\t}= \I{t}^I_{\xi,\t}(u\t,\I{u}^I_{\xi,\t})$,
\end{lst}

\noindent
where $\xi_x^S=\xi\cup\xS$ and $\t_x^u=\t\cup\xu$. A substitution $\t$
is {\em $I$-adapted} to a $\G$-assignment $\xi$ if
$\dom(\t)\sle\dom(\G)$ and, for all $x\in\dom(\t)$, $x\t\in
\I{x\G}^I_{\xi,\t}$. A pair $(\xi,\t)$ is {\em $(\G,I)$-valid},
written $\xi,\t\models_I\G$, if $\xi\models\G$ and $\t$ is $I$-adapted
to $\xi$. A term $t$ such that $\G\th t:T$ is {\em computable} if, for
all $(\G,I)$-valid pair $(\xi,\t)$, $t\t\in\I{T}^I_{\xi,\t}$. A
sub-system $\th'\,\sle\,\th$ is {\em computable} if every term typable
in it is computable.
\end{definition}

Thanks to the property satisfied by $\cN$, one can prove that the
interpretation schema defines reducibility candidates: if $\G\th t:T$
and $\xi\models\G$, then $\I{t}^I_{\xi,\t}\in\cR_T$ (see Lemma 38 in
\cite{blanqui05mscs}). Note also that $\I{t}_{\xi,\t}^I=
\I{t}_{\xi',\t'}^{I'}$ whenever $\xi$ and $\xi'$ agree on the
predicate variables free in $t$, $\t$ and $\t'$ agree on the variables
free in $t$, and $I$ and $I'$ agree on the symbols occurring in $t$.

Now, the difficult point is to define an interpretation $I$ for every
predicate symbol and to prove that every symbol $f$ is computable, \ie
$f\in\I{\tf}^I$. We define $I$ by induction on the precedence, and
simultaneously for the symbols that are in the same equivalence
class. We first give the interpretation for defined predicate symbols.


\begin{definition}[Interpretation of defined predicate symbols]
\label{def-int-def}
If every $t_i$ has a normal form $t_i^*$ and $\vt^*=\vl\s$ for some
rule $F\vl\a r\in\cR$, then $I_F(\vt,\vS)= \I{r}^I_{\xi,\s}$ with
$x\xi=S_{\ka_x}$. Otherwise, $I_F(\vt,\vS)=\SN$.
\end{definition}

Sufficient conditions of well-definedness are given in
\cite{blanqui05mscs}. Among other things, it assumes that, for every
rule $f\vl\a r$, every symbol occurring in $r$ is smaller than $f$
(see Definition \ref{def-rule}).

In order for the interpretation to be compatible with the conversion
rule, we must make sure that $\I{T}^I_{\xi,\t}=\I{T'}^I_{\xi,\t}$
whenever $T\a T'$. This property is easily verified if predicate-level
rewrite rules have critical pairs with no rule, as required in
Definition \ref{def-cond-pred} (see Lemma 65 in \cite{blanqui05mscs}).


Now, following previous works on inductive types
\cite{mendler87thesis,werner94thesis}, the interpretation of a
constant predicate symbol $C$ is defined as the least fixpoint of a
monotone function $\vphi_C$ on the complete lattice
$\cR_\tC$. Following Matthes \cite{matthes98thesis}, there are
essentially two possible definitions that we illustrate by the case of
$nat$. The {\em introduction-based} definition:

\begin{center}
$\vphi_{nat}(I)= \{t\in\SN~|~ t\a^* su\A u\in I\}$
\end{center}

\noindent
and the {\em elimination-based} definition:

\begin{center}
$\vphi_{nat}(I)= \{t\in\cT~|~ \all (\xi,\t)\, (\G,I)\mbox{-valid},\,
rec~P\t~u\t~v\t~t\in\I{Pn}^I_{\xi,\t_n^t}\}$
\end{center}

\noindent
where $\G=P:nat\A\st,u:P0,v:(n:nat)Pn\A P(sn)$. In both cases, the
monotony of $\vphi_{nat}$ is ensured by the fact that $nat$ occurs
only {\em positively} in the types of the arguments of its
constructors, a common condition for inductive types (for simple
types, we say that $X$ occurs positively in $Y\A X$ and negatively in
$X\A Y$). Indeed, Mendler proved that recursors for negative types are
not normalizing \cite{mendler87thesis}. Take for instance an inductive
type $C$ with constructor $c:(C\A nat)\A C$. Assume now that we have
$p:C\A (C\A nat)$ defined by the rule $p(cx)\ar x$. Then, by taking
$\w= [x:C](px)x$, we get the infinite reduction sequence $\w(c\w)\ab
p(c\w)(c\w)\ar \w(c\w)\ab \ldots$ We now extend the notion of positive
positions to the terms of CC (in Section \ref{sec-indrec}, we give a
more general definition for dealing with inductive-recursive types):


\begin{definition}[Positive/negative positions]
\label{def-pos}
The sets of {\em positive positions} $\pos^+(t)$ and {\em negative
positions} $\pos^-(t)$ in a term $t$ are inductively defined as
follows:

\begin{lst}{--}
\item $\pos^\d(s)= \pos^\d(x)= \pos^\d(f)= \{\vep~|~\d=+\}$,
\item $\pos^\d((x:U)V)= 1.\pos^{-\d}(U)\cup 2.\pos^\d(V)$,
\item $\pos^\d([x:U]v)= 2.\pos^\d(v)$,
\item $\pos^\d(tu)= 1.\pos^\d(t)$,
\end{lst}

\noindent
where $\vep$ is the empty word, ``.'' the concatenation,
$\d\in\{-,+\}$, $-+=-$ and $--=+$ (usual rules of signs). Moreover, if
$\le$ is an ordering, we let $\le^+=\le$ and $\le^-=\ge$.
\end{definition}


In \cite{blanqui05mscs}, we used the introduction-based approach since
this allowed us to have non-free constructors and pattern-matching on
defined symbols, which is forbidden in CIC and does not seem possible
with the elimination-based approach. For instance, in CAC, it is
possible to formalize the type $int$ of integers by simply taking the
symbols $0:int$, $s:int\A int$ and $p:int\A int$, together with the
rules:

\begin{rewc}
s~(p~x) & x\\
p~(s~x) & x\\
\end{rewc}

It is also possible to have the following rule on natural numbers:

\begin{rewc}
x\times(y+z) & (x\times y)+(x\times z)\\
\end{rewc}

To this end, we considered as constructor not only the usual
(constant) constructor symbols but any symbol $c$ whose output type is
a constant predicate symbol $C$ (perhaps applied to some
arguments). Then, to preserve the monotony of $\vphi_C$, matching
against $c$ is restricted to the arguments, called {\em accessible},
in the type of which $C$ occurs only positively. We denote by
$\acc(c)$ the set of accessible arguments of $c$. For instance, $x$ is
accessible in $sx$ since $nat$ occurs only positively in the type of
$x$. But, we also have $x$ and $y$ accessible in $x+y$ since $nat$
occurs only positively in the types of $x$ and $y$. So, $+$ can be
seen as a constructor too, whose arguments are both accessible.

With this approach, we can safely take:

\begin{center}
$\vphi_{nat}(I)= \{t\in\SN~|~ \all f, t\a^* f\vu\A \all j\in \acc(f),
u_j\in \I{U_j}^I_{\xi,\t}\}$
\end{center}

\noindent
where $f$ is any symbol of type $(\vy:\vU)nat$ and $\t=\vyu$, whenever
an appropriate assignment $\xi$ for the predicate variables of $U_j$
can be defined, which seems possible only if the condition (I6) is
satisfied (see Definition \ref{def-i6}). Here, since $nat$ has no
parameter, this condition is satisfied only if $U_j$ has no predicate
argument.

As a consequence, if $f\vt$ is computable then, for all $j\in\acc(f)$,
$t_j$ is computable (see Lemma 53 in \cite{blanqui05mscs}). This means
that, when a rule applies, the matching substitution $\s$ is
computable. This property is then used for proving the termination of
higher-order rewrite rules by using the notion of computability
closure of a rule left hand-side (see Definition 25 in
\cite{blanqui05mscs}). The computability closure is defined in such a
way that, if $r$ is in the computability closure of $f\vl$ then, for
all computable substitution $\s$, $r\s$ is computable whenever the
terms in $\vl\s$ are computable (see Theorem 67 in
\cite{blanqui05mscs}).

As for first-order rewrite rules, \ie rules with algebraic right
hand-sides and variables of first-order data type only, it is well
known since the pioneering works of Breazu-Tannen and Gallier
\cite{breazu89icalp}, and Okada \cite{okada89issac}, that their
combination with non-dependent typed $\la$-calculi preserves strong
normalization. It comes from the fact that first-order rewriting
cannot create new $\b$-redexes. This result can be extended to our
more general framework if the following two conditions are satisfied:

\begin{lst}{--}
\item Since we consider the combination of a set of first-order rewrite
rules and a set of higher-order rewrite rules, and since strong
normalization is not modular \cite{toyama87ipl}, we require
first-order rewrite rules to be non duplicating (no variable occurs
more times in a right hand-side than in a left hand-side)
\cite{rusinowitch87ipl,jouannaud97tcs}.

\item For proving that first-order rewrite rules preserve
not only strong normalization but also computability, we must make
sure that, for first-order data types, computability is equivalent to
strong normalization.
\end{lst}


In fact, we consider a slightly more general notion of first-order
data type than usual: our first-order data types can be dependent if
the dependencies are first-order data types too (\eg lists of natural
numbers of fixed length).

\begin{definition}[First-order data types]
Types equivalent to $C$ are {\em first-order data
types}\footnote{Called {\em primitive} in \cite{blanqui05mscs}.} if,
for all $D\simeq C$, $D:(\vz:\vV)\st$, $\{\vz\}\sle\Xs$ and, for all
$d:(\vx:\vT)D\vv$, $\{\vx\}\sle\Xs$, $\acc(d)=\{1,\ldots,|\vx|\}$ and
every $T_j$ is of the form $E\vw$ with $E\le C$ a first-order data
type too.
\end{definition}


\section{Abstract recursors}

From now on, we assume that the set of constant predicate symbols
$\CFB$ is divided in two disjoint sets: the set $\CFBI$ of predicate
symbols interpreted by the introduction-based method of
\cite{blanqui05mscs}, and the set $\CFBE$ of predicate symbols
interpreted by the elimination-based method of the present paper.

We now introduce an abstract notion of recursor for dealing with the
elimination-based method in a general way.

\begin{definition}[Pre-recursors]
\label{def-rec}
A {\em pre-recursor} for a symbol $C:(\vz:\vV)\st$ in $\CFBE$ is any
symbol $f\notin\cons$ such that:

\begin{lst}{--}
\item $\tf$ is of the form $(\vz:\vV)(z:C\vz)W$,
\item every predicate symbol occurring in $W$ is smaller than $C$,
\item every rule defining $f$ is of the form $f\vz(c\vt)\vu\a r$ with
$c$ constant, $\vz\in\cX$ and $\FV(r)\cap \{\vz\}=\vide$,
\end{lst}
\end{definition}

The form of a pre-recursor type may seem restrictive. However, since
termination is not established yet, we cannot consider the normal form
of a type when testing if it matches some given form. Moreover, in an
environment, every two variables whose types do not depend on each
other can be permuted without modifying the set of terms typable in
this environment (see Lemma 18 in \cite{blanqui01thesis}). So, our
results also apply on symbols whose type can be brought to this form
by various applications of this lemma.


\begin{definition}[Positivity conditions]
\label{def-pos-cond}
A pre-recursor $f:(\vz:\vV)(z:C\vz)W$ is a {\em recursor} if it
satisfies the following {\em positivity conditions}:\footnote{In
Section \ref{sec-indrec}, we give weaker conditions for dealing with
inductive-recursive types.}

\begin{lst}{--}
\item no defined predicate $F\simeq C$ occurs in $W$: $\pos(F,W)=\vide$,
\item every constant predicate $D\simeq C$ occurs only positively in $W$:
$\pos(D,W)\sle\pos^+(W)$.
\end{lst}

\noindent
A recursor $f$ of sort $s_f=\st$ (resp. $\B$) is {\em weak}
(resp. {\em strong}). We assume that every type $C\in\CFBE$ has a non
empty set $\rec(C)$ of recursors, and that $\rec(C)\cap\rec(D)=\vide$
whenever $C$ and $D$ are two distinct predicate symbols of $\CFBE$.
\end{definition}


We now define a set $\cN$ of {\em neutral terms} (see Definition
\ref{def-cand}) that is adapted to both the introduction-based and the
elimination-based approach.

\begin{definition}[Neutral terms]
\label{def-neutr}
For the set $\cN$ of {\em neutral terms} (see Definition
\ref{def-cand}), we choose the set of all terms not of the form:

\begin{lst}{--}
\item abstraction: $[x:T]u$,
\item partial application: $f\vt$ with $f$ defined by some rule $f\vl\a r$
with $|\vl|>|\vt|$,
\item constructor: $f\vt$ with $\tf=(\vy:\vU)C\vv$, $|\vt|=|\vy|$,
$C\in\CFB$, and $f$ constant whenever $C\in\CFBE$.
\end{lst}
\end{definition}

In comparison with Definition 31 in \cite{blanqui05mscs}, we just
added the restriction, in the constructor case, that $f$ is constant
if $C\in\CFBE$. This therefore changes nothing if $C\in\CFBI$.


We now define the interpretation of the equivalence class of a symbol
$C\in\CFBE$. Since we proceed by induction on the precedence for
defining the interpretation of predicate symbols, we can assume that
an interpretation for the symbols strictly smaller than $C$ is already
defined. The set of interpretations for constant predicate symbols
equivalent to $C$, ordered point-wise, is a complete lattice. We now
define the monotone function $\vphi$ on this lattice whose fixpoint
will be the interpretation for constant predicate symbols equivalent
to $C$.

\begin{definition}[Interpretation of constant predicate symbols from $\CFBE$]
\label{def-int}
If every $t_i$ has a normal form $t_i^*$ then $\vphi^I_C(\vt,\vS)$ is
the set of terms $t$ such that, for all $f\in\rec(C)$ of type
$(\vz:\vV)(z:C\vz)(\vy:\vU)V$ with $V$ not a product, and for all
$\vy\xi$ and $\vy\t$, if $\xi_\vz^\vS,\t_\vz^\vt{}_z^t\models_I
\vy:\vU$ then $f\vt^* t\vy\t\in
\I{V}_{\xi_\vz^\vS,\t_\vz^\vt{}_z^t}^I$. Otherwise,
$\vphi^I_C(\vt,\vS)=\SN$.
\end{definition}

This interpretation is well defined since, by Definition
\ref{def-rec}, every predicate symbol occurring in ${(\vy:\vU)}V$ is
smaller than $C$. Furthermore, one can easily check that $\vphi^I_C$
is stable by reduction: if $\vt\a\vt'$ then
$\vphi^I_C(\vt,\vS)=\vphi^I_C(\vt',\vS)$. We now prove that
$\vphi^I_C(\vt,\vS)$ is a reducibility candidate.


\begin{lemma}
$R=\vphi^I_C(\vt,\vS)$ is a reducibility candidate.
\end{lemma}

\begin{proof}
\begin{enumi}{R}
\item Let $t\in R$. We must prove that $t\in\SN$. Since
$\rec(C)\neq\vide$, there is at least one recursor $f$. Take
$y_i\t=y_i$ and $y_i\xi=\top_{U_i}$. We clearly have
$\xi_\vz^\vS,\t_\vz^\vt{}_z^t\models_I\vy:\vU$. Therefore, $f\vt^*
t\vy\in S=\I{V}_{\xi_\vz^\vS,\t_\vz^\vt{}_z^t}^I$. Now, since $S$
satisfies (R1), $f\vt^*t\vy\in\SN$ and $t\in\SN$.

\item Let $t\in R$ and $t'\in\,\a\!\!(t)$. We must prove that $t'\in
R$, hence that $f\vt^*t'\vy\t\in
S=\I{V}_{\xi_\vz^\vS,\t_\vz^\vt{}_z^t}^I$. This follows from the fact
that $f\vt^*t\vy\t\in S$ (since $t\in R$) and $S$ satisfies (R2).

\item Let $t$ be a neutral term such that $\a\!\!(t)\sle R$. We must
prove that $t\in R$, hence that $u=f\vt^* t\vy\t\in
S=\I{V}_{\xi_\vz^\vS,\t_\vz^\vt{}_z^t}^I$. Since $u$ is neutral and
$S$ satisfies (R3), it suffices to prove that $\a\!\!(u)\sle S$. Since
$\vy\t\in\SN$ by (R1), we proceed by induction on $\vy\t$ with $\a$ as
well-founded ordering. The only difficult case could be when $u$ is
head-reducible, but this is not possible since $t$ is neutral.
\end{enumi}
\end{proof}


The fact that $\vphi$ is monotone, hence has a least fixpoint, follows
from the positivity conditions.

\begin{lemma}
\label{lem-I-mon}
Let $I\le_f I'$ iff $I_f\le I'_f$ and, for all $g\neq f$,
$I_g=I'_g$. If $I\le_f I'$, $\pos(f,t)\sle\pos^\d(t)$, $\G\th t:T$ and
$\xi\models\G$ then $\I{t}^I_{\xi,\t}\le^\d \I{t}^{I'}_{\xi,\t}$.
\end{lemma}

\begin{proof}
By induction on $t$.

\begin{lst}{--}
\item $\I{s}^I_{\xi,\t}= \top_s= \I{s}^{I'}_{\xi,\t}$.

\item $\I{x}^I_{\xi,\t}= x\xi= \I{x}^{I'}_{\xi,\t}$.

\item Let $R=\I{g\vt}^I_{\xi,\t}$ and
$R'=\I{g\vt}^{I'}_{\xi,\t}$. $R=I_g(\vt\t,\vS)$ with
$\vS=\I{\vt}^I_{\xi,\t}$. $R'=I'_g(\vt\t,\vS')$ with
$\vS'=\I{\vt}^{I'}_{\xi,\t}$. Since $\pos(f,\vt)=\vide$,
$\vS=\vS'$. Now, if $f=g$ then $R\le R'$ and $\d=+$
necessarily. Otherwise, $R=R'$.

\item Let $R=\I{(x:U)V}^I_{\xi,\t}$ and
$R'=\I{(x:U)V}^{I'}_{\xi,\t}$. $R=\{t\in\cT~|~ \all u\in
\I{U}^I_{\xi,\t}, \all S\in\cR_U, tu\in
\I{V}^I_{\xi_x^S,\t_x^u}\}$. $R'= \{t\in\cT~|~ \all u\in
\I{U}^{I'}_{\xi,\t}, \all S\in\cR_U, tu\in
\I{V}^I_{{\xi'}_x^S,\t_x^u}\}$. Since $\pos^\d((x:U)V)= 1.\pos^{-\d}(U)\cup
2.\pos^\d(V)$, $\pos(f,U)$ $\sle\pos^{-\d}(U)$ and
$\pos(f,V)\sle\pos^\d(V)$. Therefore, by induction hypothesis,
$\I{U}^I_{\xi,\t}\le^{-\d} \I{U}^{I'}_{\xi,\t}$ and
$\I{V}^I_{\xi_x^S,\t_x^u}\le^\d \I{V}^I_{{\xi'}_x^S,\t_x^u}$. So,
$R\le^\d R'$. Indeed, if $\d=+$, $t\in R$ and $u\in
\I{U}^{I'}_{\xi,\t}\sle \I{U}^I_{\xi,\t}$ then $tu\in
\I{V}^I_{\xi_x^S,\t_x^u}\sle \I{V}^I_{{\xi'}_x^S,\t_x^u}$ and $t\in
R'$. If $\d=-$, $t\in R'$ and $u\in \I{U}^I_{\xi,\t}\sle
\I{U}^{I'}_{\xi,\t}$ then $tu\in \I{V}^I_{{\xi'}_x^S,\t_x^u}\sle
\I{V}^I_{\xi_x^S,\t_x^u}$ and $t\in R$.

\item Let $R=\I{[x:U]v}^I_{\xi,\t}$ and
$R'=\I{[x:U]v}^{I'}_{\xi,\t}$. $R$ and $R'$ have the same domain
$\cT\times\cR_U$ and the same codomain
$\cR_V$. $R(u,S)=\I{v}^I_{\xi_x^S,\t_x^u}$ and
$R'(u,S)=\I{v}^I_{{\xi'}_x^S,\t_x^u}$. Since $\pos^\d([x:U]v)=
2.\pos^\d(v)$, $\pos(f,v)\sle\pos^\d(v)$. Therefore, by induction
hypothesis, $R(u,S)\le^\d R'(u,S)$ and $R\le^\d R'$.

\item Let $R=\I{tu}^I_{\xi,\t}$ and $R'=\I{tu}^{I'}_{\xi,\t}$.
$R=\I{t}^I_{\xi,\t}(u\t,S)$ with
$S=\I{u}^I_{\xi,\t}$. $R'=\I{t}^{I'}_{\xi,\t}(u\t,S')$ with
$S'=\I{u}^{I'}_{\xi,\t}$. Since $\pos^\d(tu)= 1.\pos^\d(t)$,
$\pos(f,t)\sle\pos^\d(t)$ and $\pos(f,u)=\vide$. Therefore, $S=S'$
and, by induction hypothesis, $\I{t}^I_{\xi,\t}\le^\d
\I{t}^{I'}_{\xi,\t}$. So, $R\le^\d R'$.
\end{lst}
\end{proof}


\begin{lemma}
$\vphi$ is monotone.
\end{lemma}

\begin{proof}
Let $I\le J$. We must prove that, for all $C$, $\vt$, $\vS$,
$\vphi_C^I(\vt,\vS)\sle \vphi_C^J(\vt,\vS)$. If some $t_i$ has no
normal form then $\vphi_C^I(\vt,\vS)= \vphi_C^J(\vt,\vS)= \SN$. Assume
now that every $t_i$ has a normal form $t_i^*$. Let $t\in
\vphi_C^I(\vt,\vS)$, $f\in\rec(C)$ with
$\tf=(\vz:\vV)(z:C\vz)(\vy:\vU)V$, $\vy\xi$ and $\vy\t$ such that
$\xi_\vz^\vS,\t_\vz^\vt{}_z^t\models_J \vy:\vU$. We must prove that
$f\vt^* t\vy\t\in \I{V}^J_{\xi_\vz^\vS,\t_\vz^\vt{}_z^t}$.
$\xi_\vz^\vS,\t_\vz^\vt{}_z^t\models_J \vy:\vU$ means that $\vy\t\in
\I{\vU}^J_{\xi_\vz^\vS,\t_\vz^\vt{}_z^t}$.

Let $W=(\vy:\vU)V$. By assumption, for every $D\simeq C$,
$\pos(D,W)\sle\pos^+(W)$. Thus, $\pos(D,\vU)\sle\pos^-(\vU)$ and
$\pos(D,V)\sle\pos^+(V)$. Hence, by Lemma \ref{lem-I-mon},
$\xi_\vz^\vS,\t_\vz^\vt{}_z^t\models_I \vy:\vU$ and
$\I{V}^I_{\xi_\vz^\vS,\t_\vz^\vt{}_z^t}\sle
\I{V}^J_{\xi_\vz^\vS,\t_\vz^\vt{}_z^t}$. Thus, $f\vt^*
t\vy\t\in \I{V}^J_{\xi_\vz^\vS,\t_\vz^\vt{}_z^t}$.
\end{proof}


\section{Admissible recursors}
\label{sec-admis}

Now, for getting termination of $\b\cup\cR$, we need to prove that
every symbol $f$ is computable, \ie $f\in\I\tf$. To this end, we give
general conditions on recursors. We focus on what is new and refer the
reader to \cite{blanqui05mscs} for the other cases. After Lemma
\ref{lem-ind-prec}, we know that we can proceed by induction on the
precedence for proving the computability of well-typed terms. So, when
defining conditions on a symbol $f$, we can always assume
w.l.o.g. that $\th_f^<$ is computable, \ie terms with symbols strictly
smaller than $f$ are computable (see Definition
\ref{def-schema-int}). In particular, every subterm of $\tf$ is
computable (see Corollary \ref{cor-typ}).


\begin{definition}[Admissible recursors]
\label{def-rec-admis}
Let $C:(\vz:\vV)\st$ be a constant predicate symbol such that
$\rec(C)\neq\vide$. We assume that every symbol $c:(\vx:\vT)C\vv$ is
equipped with a set $\acc(c)\sle\{1,\ldots,|\vx|\}$ of {\em accessible
arguments}. A {\em constructor} of $C$ is any constant symbol
$c:(\vx:\vT)C\vv$.

The set $\rec(C)$ is {\em complete w.r.t. accessibility} if, for all
constructor $c:(\vx:\vT)C\vv$, $j\in\acc(c)$, $\vx\eta$ and $\vx\s$,
if $\eta\models\G_c$, $\vv\s\in\SN$ and $c\vx\s\in
\I{C\vv}_{\eta,\s}$ then $x_j\s\in \I{T_j}_{\eta,\s}$.

A recursor $f:(\vz:\vV)(z:C\vz)(\vy:\vU)V$ is {\em head-computable
w.r.t.} a constructor $c:(\vx:\vT)C\vv$ if, whenever $\th_f^<$ is
computable, for all $\vx\eta$, $\vx\s$, $\vy\xi$, $\vy\t$,
$\vS=\I{\vv}_{\eta,\s}$ such that $\eta,\s\models\G_c$ and
$\xi_\vz^\vS,\t_\vz^{\vv\s}{}_z^{c\vx\s}\models\vy:\vU$, every
head-reduct of $f\vv\s(c\vx\s)\vy\t$ belongs to
$\I{V}_{\xi_\vz^\vS,\t_\vz^{\vv\s}{}_z^{c\vx\s}}$. A recursor is {\em
head-computable} if it is head-computable w.r.t. every constructor of
$C$. $\rec(C)$ is {\em head-computable} if all its recursors are
head-computable.

$\rec(C)$ is {\em admissible} if it is head-computable and complete
w.r.t. accessibility.
\end{definition}


Completeness w.r.t. accessibility exactly insures that, if $c\vt$ is
computable then, for all $j\in\acc(c)$, $t_j$ is computable (Lemma 53
in \cite{blanqui05mscs}), hence that non-recursor higher-order symbols
are computable (see Lemma 68 in \cite{blanqui05mscs}). We now prove
that the elimination-based interpretation of first-order data types is
$\SN$, hence that first-order symbols are computable (see Lemma 63 in
\cite{blanqui05mscs}).

\begin{lemma}
\label{lem-prim}
If $C$ is a first-order data type and $\rec(C)$ is head-computable
then $I_C(\vt,\vS)=\SN$.
\end{lemma}

\begin{proof}
First note that $S_i=\vide$ since $\{\vz\}\sle\Xs$. So, we do not
write $\vS$ in the following. By definition, for all $\vt$,
$I_C(\vt)\sle\SN$. We now prove that, if $t\in\SN$ then, for all
$\vt$, $t\in I_C(\vt)$, by induction on $t$ with $\a\cup\,\tgt$ as
well-founded ordering. If some $t_i$ has no normal form then $t\in
I_C(\vt)=\SN$. Assume now that every $t_i$ has a normal form
$t_i^*$. Let $f:(z:C)(\vy:\vU)V$ be a recursor of $C$, $\vy\xi$,
$\vy\t$ and $\s=\t_\vz^\vt{}_z^t$ such that $\xi,\s\models
\vy:\vU$. We must prove that $v=f\vt^*t\vy\t\in
S=\I{V}_{\xi,\s}$. Since $v$ is neutral, it suffices to prove that
$\a\!\!(v)\sle S$. We proceed by induction on $t\vy\t$ with $\a$ as
well-founded ordering ($\vy\t\in\SN$ by R1). If the reduction takes
place in $t\vy\t$, we can conclude by induction hypothesis. Assume now
that $v'$ is a head-reduct of $v$. By assumption on recursors, $t$ is
of the form $c\vu$ with $c:(\vx:\vT)C\vv$. Let $\g=\vxu$. Since $C$ is
a first-order data type, every $u_j$ is accessible and every $T_j$ is
of the form $D\vw$ with $D$ a first-order data type too. Thus, by
induction hypothesis, for all $j$, $u_j\in I_D(\vw\g)$. Therefore,
$\vide,\g\models\G_c$ and $v'\in S$ since $\xi,\s\models \vy:\vU$ and
recursors are assumed to be head-computable.
\end{proof}


\begin{lemma}
\label{lem-head-comp}
Head-computable recursors are computable.
\end{lemma}

\begin{proof}
Let $f:(\vz:\vV)(z:C\vz)(\vy:\vU)V$ be a recursor and assume that
$\xi,\t\models\G_f$. We must prove that $v=f\vz\t z\t\vy\t\in
S=\I{V}_{\xi,\t}$. Since $v$ is neutral, it suffices to prove that
$\a\!\!(v)\sle S$. We proceed by induction on $\vz\t z\t\vy\t$ with
$\a$ as well-founded ordering ($\vz\t z\t\vy\t\in\SN$ by R1). If the
reduction takes place in $\vz\t z\t\vy\t$, we conclude by induction
hypothesis. Assume now that we have a head-reduct $v'$. By definition
of recursors (see Definition \ref{def-rec}), $z\t$ is of the form
$c\vu$ with $c:(\vx:\vT)C\vv$, and $v'$ is also a head-reduct of $v_0=
f(\vz\t)^*z\t\vy\t$. Since $\xi,\t\models\G_f$, we have $z\t=c\vu\in
\I{C\vz}_{\xi,\t}= I_C(\vz\t,\vz\xi)$. Therefore, by definition of
$I_C$, $v_0\in S$ and, by (R2), $v'\in S$.
\end{proof}


\begin{lemma}[Computability]
\label{lem-comp}
For all $g$, if $\th_g^<$ is computable then $\th_g$ is computable. 
\end{lemma}

\begin{proof}
We prove that, if $\G\th_g t:T$ and $\eta,\s\models\G$ then
$t\s\in\I{T}_{\eta,\s}$, by induction on $\G\th_g t:T$. We only detail
the (symb) case. The other cases are detailed in Lemma 66 in
\cite{blanqui05mscs}. So, assume that $\th_g f:\tf$. If $f<g$ then, by
Lemma \ref{lem-ind-prec}, $\th_g^< f:\tf$ and $f$ is computable since
$\th_g^<$ is assumed to be computable. Otherwise, $f\simeq g$ and
$\th_f^<=\th_g^<$. If $f$ is a recursor then we can conclude by Lemma
\ref{lem-head-comp}. So, assume that $f$ is not a recursor and that
$\tf=(\vx:\vT)U$ with $U$ distinct from a product. By Definition
\ref{def-schema-int}, $f$ is computable iff, for all $\G_f$-valid pair
$(\eta,\s)$, $t=f\vx\s\in R=\I{U}_{\eta,\s}$.

If $t$ is neutral then, by definition \ref{def-cand}, it suffices to
prove that $\a\!\!(t)\sle R$, which follows from Lemmas 63 and 68 in
\cite{blanqui05mscs}. Assume now that $t$ is not neutral. Then, $U=C\vv$
with $C\in\CFB$, and $R=I_C(\vv\s,\vS)$ with
$\vS=\I{\vv}_{\eta,\s}$. If $C\in\CFBI$ then, again, it follows from
Lemmas 63 and 68 in \cite{blanqui05mscs}. Otherwise, $C\in\CFBE$ and,
by Definition \ref{def-rec}, $f$ is constant.

By Corollary \ref{cor-typ}, $\th_f^<\tf:s_f$. Since, by assumption,
$\th_f^<$ is computable, by (R1), $\vv\s\in\SN$. So, let
$g:(\vz:\vV)(z:C\vz)(\vy:\vU)V$ be a recursor of $C$, $\vy\xi$ and
$\vy\t$ such that
$\xi_\vz^\vS,\t_\vz^{\vv\s}{}_z^{f\vx\s}\models\vy:\vU$. We must prove
that $v=g(\vv\s)^*(f\vx\s)\vy\t\in
S=\I{V}_{\xi_\vz^\vS,\t_\vz^{\vv\s}{}_z^{f\vx\s}}$. Since $v$ is
neutral, it suffices to prove that $\a\!\!(v)\sle S$. By (R1),
$\vx\s\vy\t\in\SN$. So, we can proceed by induction on $\vx\s\vy\t$
with $\a$ as well-founded ordering. No reduction can take place at the
top of $f\vx\s$ since $f$ is constant. In the case of a reduction in
$\vx\s\vy\t$, we conclude by induction hypothesis. Finally, in the
case of a head-reduction, we conclude by head-computability of $g$.
\end{proof}


We can now state our main result:

\begin{theorem}[Strong normalization]
\label{thm-admis}
$\b\cup\cR$ preserves typing and is strongly normalizing if:

\begin{lst}{--}
\item $\b\cup\cR$ is confluent\footnote{Again, this is the case if,
for instance, $\cR$ is confluent and left-linear \cite{muller92ipl}.}
(if there are predicate-level rules),
\item rewrite rules are well-typed,
\item every constant predicate symbol $C\in\CFBE$ is equipped with an
admissible set $\rec(C)$ of recursors,
\item strong recursors and non-recursor symbols satisfy the
conditions given in Definition 29 in \cite{blanqui05mscs}.
\end{lst}
\end{theorem}

\begin{proof}
After Lemma \ref{lem-ind-prec}, we can proceed by induction on the
precedence. Hence, by Lemma \ref{lem-comp}, every well-typed term is
computable. Let $t$ be a term such that $\G\th t:T$. With $x\t=x$ and
$x\xi=\top_{x\G}$, we clearly have $\xi,\t\models\G$ since, by Lemma
33 in \cite{blanqui05mscs}, variables are elements of every
candidate. Thus, by (R1), $t\in\SN$.
\end{proof}

As an application example of this theorem, we prove just below the
admissibility of a large class of recursors for strictly positive
types, from which Coq's recursors \cite{coqv80} can be easily derived
(see Section \ref{sec-cic}). Before that, let us remark that the
condition I6 and the safeness condition described in the introduction
(Definitions \ref{def-i6} and \ref{def-safe} respectively) are not
necessary anymore for weak recursors. On the other hand, the safeness
condition is still necessary for non-recursor symbols and strong
recursors on types like $\JMeq$.


\begin{definition}[Canonical recursors for strictly positive types]
\label{def-can-rec}
Let $C:(\vz:\vV)\st$ and $\vc$ be {\em strictly positive} constructors
of $C$, that is, if $c_i$ is of type $(\vx:\vT)C\vv$ then either no
type equivalent to $C$ occurs in $T_j$ or $T_j$ is of the form
$(\vec\alpha:\vW)C\vw$ with no type equivalent to $C$ in $\vW$. The
{\em parameters} of $C$ are the biggest sequence $\vq$ such that
$C:(\vq:\vQ)(\vz:\vV)\st$ and each $c_i$ is of type
$(\vq:\vQ)(\vx:\vT)C\vq\vv$ with $T_j= (\vec\alpha:\vW)C\vq\vw$ if $C$
occurs in $T_j$.

The {\em canonical weak recursor} of $C$ w.r.t. $\vc$ is
$rec^\st_{\vc}:(\vq:\vQ)(\vz:\vV)(z:C\vq\vz){(P:(\vz:\vV)C\vq\vz\A\st)}$
${(\vy:\vU)}P\vz z$ with $U_i= (\vx:\vT)(\vx':\vT')P\vv(c_i\vq\vx)$,
$T_j'= {(\vec\alpha:\vW)}P\vw(x_j\vec\alpha)$ if $T_j=
{(\vec\alpha:\vW)}C\vq\vw$, and $T_j'=T_j$ otherwise, defined by the
rules $rec^\st_{\vc}\vq\vz(c_i\vq'\vx)P\vy\a y_i\vx\vt'$ where
$\vq,\vz,\vq',\vx,P,\vy$ are variables, $t_j'=
{[\vec\alpha:\vW]}(rec^\st_{\vc}\vq\vw(x_j\vec\alpha)P\vy)$ if $T_j=
{(\vec\alpha:\vW)}C\vq\vw$, and $t_j'=x_j$ otherwise.\footnote{We
could erase the useless arguments $t_j'=x_j$ when $T_j'=T_j$ as it is
done in CIC.}

The {\em canonical strong recursor}\footnote{Strong recursors cannot
be defined exactly like weak recursors by simply taking
$P:(\vz:\vV)C\vq\vz\A\B$ since $(\vz:\vV)C\vq\vz\A\B$ is not typable
in CC. They must be defined for each $P$. That is why Werner
considered a slightly more general PTS in \cite{werner94thesis}.} of
$C$ w.r.t. $\vc$ and $P=[\vz:\vV][z:C\vq\vz]Q$ is
$rec^P_{\vc}:(\vq:\vQ)(\vz:\vV)$ $(z:C\vq\vz)(\vy:\vU)Q$ with $U_i=
(\vx:\vT)(\vx':\vT')Q\{\vz\to\vv,z\to c_i\vq\vx\}$, $T_j'=
{(\vec\alpha:\vW)}Q\{\vz\to\vw,z\to x_j\vec\alpha\}$ if $T_j=
{(\vec\alpha:\vW)}C\vq\vw$, and $T_j'=T_j$ otherwise, defined by the
rules $rec^P_{\vc}\vq\vz(c_i\vq'\vx)\vy\a y_i\vx\vt'$ where
$\vq,\vz,\vq',\vx,\vy$ are variables, $t_j'=
{[\vec\alpha:\vW]}(rec^P_{\vc}\vq\vw(x_j\vec\alpha)\vy)$ if $T_j=
{(\vec\alpha:\vW)}C\vq\vw$, and $t_j'=x_j$ otherwise.
\end{definition}


\begin{lemma}
\label{lem-sr}
The rules defining canonical recursors preserve typing.
\end{lemma}

\begin{proof}
For the rule $rec^\st_{\vc}\vq\vz(c_i\vq'\vx)P\vy\a y_i\vx\vt'$, take
$\G=\vq:\vQ,\vx:\vT,P:(\vz:\vV)C\vq\vz\A\st,\vy:\vU$ and
$\r=\{\vz\to\vv,\vq'\to\vq\}$. We prove the conditions required in
Section \ref{sec-cac}:

\begin{lst}{--}
\item One can easily check that $\G\th y_i\vx\vt':P\vv(c_i\vq\vx)$.
\item Assume now that
$\D\th (rec^\st_{\vc}\vq\vz(c_i\vq'\vx)P\vy)\s:T$. We must prove that
$\s:\G\leadsto\D$ and $\s\ad\r\s$. Both properties follow by inversion
of the typing judgment and confluence.
\end{lst}

\noindent
The proof is about the same for strong recursors.
\end{proof}


\begin{lemma}
\label{lem-comp-acc}
The set of canonical recursors is complete
w.r.t. accessibility.\footnote{In \cite{werner94thesis} (Lemma 4.35),
Werner proves a similar result.}
\end{lemma}

\begin{proof}
Let $c=c_i:(\vq:\vQ)(\vx:\vT)C\vq\vv$ be a constructor of
$C:(\vq:\vQ)(\vz:\vV)\st$, $\vq\eta$, $\vx\eta$, $\vq\s$ and $\vx\s$
such that $\vq\s\vv\s\in\SN$ and $c\vq\s\vx\s\in
\I{C\vq\vv}_{\eta,\s}= I_C(\vq\s\vv\s,\vq\eta\I{\vv}_{\eta,\s})$. Let
$\va=\vq\vx$ and $\vA=\vQ\vT$. We must prove that, for all $j$,
$a_j\s\in \I{A_j}_{\eta,\s}$. For the sake of simplicity, we assume
that weak and strong recursors have the same syntax. Since
$\vq\s\vv\s$ have normal forms, it suffices to find $P$ and $u$ such
that $rec_c\vq\vv(c\va)Pu\a u\vx\vt'\ab^* a_j$. Take
$P=[\vz:\vV][z:C\vq\vz]A_j$ and $u=[\vx:\vT][\vx':\vT']a_j$.
\end{proof}


\begin{lemma}
\label{lem-head}
Canonical recursors are head-computable.
\end{lemma}

\begin{proof}
Let
$f=rec^\st:(\vq:\vQ)(\vz:\vV)(z:C\vq\vz)(P:(\vz:\vV)C\vq\vz\A\st)(\vy:\vU)P\vz
z$ be the canonical weak recursor w.r.t. $\vc$,
$T=(\vz:\vV)C\vq\vz\A\st$, $c=c_i:(\vq:\vQ)(\vx:\vT)C\vq\vv$,
$\vq\eta$, $\vq\s$, $\vx\eta$, $\vx\s$, $P\xi$, $P\t$, $\vy\xi$,
$\vy\t$, $\vR=\I{\vv}_{\eta,\s}$, $\xi'=\xi_\vz^\vR$ and
$\t'=\t_\vz^{\vv\s}{}_z^{c\vx\s}$, and assume that $\th_f^<$ is
computable, $\eta,\s\models\G_c$ and $\eta\xi',\s\t'\models
P:T,\vy:\vU$. We must prove that $y_i\t\vx\s\vt'\s\t\in \I{P\vz
z}_{\xi',\t'}$.

We have $y_i\t\in \I{U_i}_{\xi',\t'}$, $U_i=
(\vx:\vT)(\vx':\vT')P\vv(c\vq\vx)$ and $x_j\s\in \I{T_j}_{\eta,\s}=
\I{T_j}_{\eta\xi',\s\t'}$. We prove that $t'_j\s\t\in
\I{T_j'}_{\eta\xi',\s\t'}$. If $T_j'=T_j$ then $t_j'\s\t=x_j\s$ and we
are done. Otherwise, $T_j= (\vec\alpha:\vW)C\vq\vw$, $T_j'=
(\vec\alpha:\vW)P\vw(x_j\vec\alpha)$ and $t_j'=
[\vec\alpha:\vW]f\vq\vw(x_j\vec\alpha)P\vy$. Let $\vec\alpha\zeta$ and
$\vec\alpha\g$ such that $\eta\xi'\zeta,\s\t'\g\models
\vec\alpha:\vW$. Let $t=x_j\s\vec\alpha\g$. We must prove that
$v=f\vq\s\vw\s\g tP\t\vy\t\in
S=\I{P\vw(x_j\vec\alpha)}_{\eta\xi'\zeta,\s\t'\g}$. Since $v$ is
neutral, it suffices to prove that $\a\!\!(v)\sle S$.

By (R1), we have $\vq\s tP\t\vy\t\in\SN$. Since $\th_f^<$ is
computable and $\vw$ is a subterm of $\tf$, by (R1), we also have
$\vw\s\g\in\SN$. Thus, we can proceed by induction on $\vq\s\vw\s\g
tP\t\vy\t\in\SN$ with $\a$ as well-founded ordering. In the case of a
reduction in $\vq\s\vw\s\g tP\t\vy\t$, we conclude by induction
hypothesis. Assume now that we have a head-reduct $v'$. By definition
of recursors, $v'$ is also a head-reduct of
$v_0=f(\vq\s)^*(\vw\s\g)^*t P\t\vy\t$ where $(\vq\s)^*(\vw\s\g)^*$ are
the normal forms of $\vq\s\vw\s\g$. If $v_0\in S$ then, by (R2),
$v'\in S$. So, let us prove that $v_0\in S$.

By candidate substitution (Lemma 40 in \cite{blanqui05mscs}),
$S=\I{P\vz z}_{\xi_\vz^\vS,\t_\vz^{\vw\s\g}{}_z^t}$ with
$\vS=\I{\vw}_{\eta\xi'\zeta,\s\t'\g}= \I{\vw}_{\eta\xi\zeta,\s\t\g}$
for $\FV(\vw)\sle\{\vq,P,\vx,\vec\alpha\}$. Since $x_j\s\in
\I{T_j}_{\eta\xi',\s\t'}$ and $\eta\xi'\zeta,\s\t'\g\models
\vec\alpha:\vW$, $t\in \I{C\vq\vw}_{\eta\xi'\zeta,\s\t'\g}=
I_C(\vq\s\vw\s\g,\vq\xi\vS)$. Since $\eta\xi',\s\t'\models
P:T,\vy:\vU$ and $\FV(T\vU)\sle\{\vq,P\}$, we have
$\eta\xi,\s\t\models P:T,\vy:\vU$ and
$\eta\xi_\vz^\vS,\s\t_\vz^{\vw\s\g}{}_z^t\models
P:T,\vy:\vU$. Therefore, $v_0\in S$.

The proof is about the same for strong recursors.
\end{proof}


\section{Application to CIC}
\label{sec-cic}

It follows that CAC subsumes CIC almost completely. However, Theorem
\ref{thm-admis} cannot be applied to CIC directly since CIC and CAC do
not have the same syntax and the same typing rules. So, we define a
sub-system of CIC, called CIC$^-$, whose terms can be translated into
a CAC satisfying the conditions of Theorem \ref{thm-admis}.


The $\io$-reduction of CIC introduces many $\b$-redexes and the
recursive calls on $Elim$ are made on bound variables which are later
instantiated by structurally smaller terms. Instead, we consider the
relation $\abip$ where one step of $\aip$ corresponds to a
$\io$-reduction followed by as many $\b$-reductions as necessary for
erasing the $\b$-redexes introduced by the $\io$-reduction. This is
this reduction relation which is actually implemented in the Coq
system \cite{coqv80}. Moreover, we conjecture that the strong
normalization of $\abip$ implies the strong normalization of $\abi$.

\begin{definition}[$\io'$-reduction]
\label{def-iotap}
The {\em $\io'$-reduction} is the reduction relation defined by the
rule:

\begin{center}
$Elim(I,Q,\vx,Constr(i,I')\,\vz)\{\vf\} ~\aip~ \D'[I,X,C_i,f_i,Q,\vf,\vz]$
\end{center}

\noindent
where $I=Ind(X:A)\{\vC\}$ and $\D'[I,X,C,f,Q,\vf,\vz]$ is defined as
follows:

\begin{lst}{--}
\item $\D'[I,X,X\vm,f,Q,\vf,\vide]= f$
\item $\D'[I,X,(z:B)D,f,Q,\vf,z\vz]= \D'[I,X,D,fz,Q,\vz]$ if $X\notin\FV(B)$
\item $\D'[I,X,(z:B)D,f,Q,\vf,z\vz]= \D'[I,X,D,fz\,[\vy:\vD]
Elim(I,Q,\vq,z\vy),Q,\vz]$ if $B=(\vy:\vD) X\vq$
\end{lst}
\end{definition}


We now define the sub-system of CIC (see Figure~\ref{fig-th-cicm})
that we are going to consider:

\begin{definition}[CIC$^-$]
\label{def-cicm}
\begin{lst}{\bu}
\item We exclude any use of the sort $\triangle$ in order to stay in
the Calculus of Constructions.

\item In the rule (conv), instead of requiring $T\aa_{\b\eta\io}^*
T'$, we require $T\aa_{\b\io'}^* T'$ which is equivalent to $T
\ad_{\b\io'} T'$ since $\abip$ is confluent (orthogonal CRS
\cite{oostrom94thesis}).

\item In the rule (Ind), we require $I$ to be in normal form w.r.t.
$\abip$ (set $\NF$) and to be typable in the empty environment since,
in CAC, the types of symbols must be typable in the empty
environment. This is not a real restriction since any type
$I=Ind{(X:A)}\{\vC\}$ typable in an environment $\G=\vy:\vU$ can be
replaced by a type $I'=Ind(X':A')\{\vC'\}$ typable in the empty
environment. It suffices to take $A'=(\vy:\vU)A$,
$C_i'=(\vy:\vU)C_i\{X\to X'\vy\}$ and to replace $I$ by $I'\vy$ and
$Constr(i,I)$ by $Constr(i,I')\vy$. Furthermore, we adapt the
definition of {\em small} constructor type accordingly. A constructor
type $C$ of an inductive type $I=Ind(X:A)\{\vC\}$ with
$A=(\vx:\vA)\st$ is {\em small} if it is of the form
$(\vx':\vA')(\vz:\vB)X\vm$ with $\vx':\vA'$ a sub-sequence of
$\vx:\vA$ and $\{\vz\}\cap\XB=\vide$.

\item In the rule ($\st$-Elim), we require $Q$ to be typable in the
empty environment, and add explicit typing judgments for $T_i$ and
$I$. Again, it is not a real restriction since we can always replace
an environment by additional abstractions.

\item In the rule ($\B$-Elim), instead of requiring
$\th Q:(\vx:\vA)I\vx\A\B$, which is not possible in CC, we require $Q$
to be of the form $[\vx:\vA][y:I\vx]K$ with $\vx:\vA,y:I\vx\th K:\B$
(this just requires some $\eta$-expansions) and $f_i$ to be of type
$T_i=\D'\{I,X,C_i,\vx y,K,Constr(i,I)\}$ where $\D'\{I,X,C,\vx
y,K,c\}$ is defined as follows:

\begin{lst}{--}
\item $\D'\{I,X,X\vm,\vx y,K,c\}= K\{\vx\to\vm,y\to c\}$,
\item $\D'\{I,X,(z:B)D,\vx y,K,c\}=\\
(z:B\XI)((\vy:\vD)K\{\vx\to\!\vq,y\to\!z\vy\})\A \D'\{I,X,D,\vx
y,K,cz\}$ if $B=(\vy:\vD)X\vq$.
\end{lst}

Moreover, we require $Q$ to be in normal form and $T_i$ to be typable.
We also take $\G\th Elim(I,Q,\va,c)$ $\{\vf\}: K\{\vx\to\va,y\to c\}$
instead of $\G\th Elim(I,Q,\va,c)\{\vf\}: Q\va c$. Finally, we require
$I$ to be safe (see Definition \ref{def-safe}): if $A=(\vx:\vA)\st$
and $C_i=(\vz:\vB)X\vm$ then:

\begin{lst}{--}
\item for all $x_i\in\XB$, $m_i\in\XB$,
\item for all $x_i,x_j\in\XB$ with $i\neq j$, $m_i\neq m_j$.
\end{lst}
\end{lst}
\end{definition}


\begin{figure}[ht]
\begin{center}
\caption{Typing rules of CIC\,$^-$\label{fig-th-cicm}}
\begin{tabular}{r@{~}c}
\\(Ind) & $\cfrac{
\begin{array}{c}
A=(\vx:\vA)\st \quad \th A:\B \quad \all i,\,X:A\th C_i:\st\\
I=Ind(X:A)\{\vC\}\in\NF \mbox{ is strictly positive}\\
\end{array}}
{\th I:A}$\\

\\(Constr) & $\cfrac{I=Ind(X:A)\{\vC\} \quad \G\th I:T}
{\G\th Constr(i,I):C_i\XI}$\\

\\($\st$-Elim) & $\cfrac{
\begin{array}{c}
A=(\vx:\vA)\st \quad I=Ind(X:A)\{\vC\}\quad \G\th I:T \quad
\th Q:(\vx:\vA)I\vx\A\st\\
T_i=\D\{I,X,C_i,Q,Constr(i,I)\} \quad \th T_i:\st\\
\all j,\, \G\th a_j:A_j\vxa \quad \G\th c:I\va \quad \all i,\,\G\th f_i:T_i\\
\end{array}}
{\G\th Elim(I,Q,\va,c)\{\vf\}:Q\va c}$\\

\\($\B$-Elim) & $\cfrac{
\begin{array}{c}
A=(\vx:\vA)\st \quad I=Ind(X:A)\{\vC\} \mbox{ is small and safe}\\
Q=[\vx:\vA][y:I\vx]K\in\NF \quad \vx:\vA,y:I\vx\th K:\B\\
T_i= \D'\{I,X,C_i,\vx y,K,Constr(i,I)\} \quad \th T_i:\B\\
\all j,\, \G\th a_j:A_j\vxa \quad \G\th c:I\va \quad \all i,\,\G\th f_i:T_i\\
\end{array}}
{\G\th Elim(I,Q,\va,c)\{\vf\}:K\{\vx\to\va,y\to c\}}$\\

\\(Conv) & $\cfrac{\G\th t:T \quad T \aa_{\b\io'}^* T' \quad \G\th
  T':s} {\G\th t:T'}$\\
\end{tabular}
\end{center}
\end{figure}


We now show that CIC$^-$ can be translated into a CAC satisfying the
conditions of Theorem \ref{thm-admis}.

\begin{definition}[Translation]
\label{def-trans}
We define $\ps{t}$ on well-typed terms, by induction on $\G\th t:T$:

\begin{lst}{\bu}
\item If $I=Ind(X:A)\{\vC\}$ then $\ps{I}=Ind_I$ where $Ind_I$
is a symbol of type $\ps{A}$.

\item $\ps{Constr(i,I)}=Constr^I_i$ where $Constr^I_i$ is a
symbol of type $\ps{C_i\{X\to I\}}$.

\item If $Q$ is not of the form $[\vx:\vA][y:I\vx](\vy:\vU)\st$
then $\ps{Elim(I,Q,\va,c)\{\vf\}}= \we\ps{Q}\ps\va\ps{c}\ps\vf$ where
$\we$ is a symbol of type ${(Q:(\vx:\ps\vA)\ps{I}\vx\A\st)}
{(\vx:\ps\vA)}{(y:\ps{I}\vx)}{(\vf:\ps\vT)}\ps{Q}\vx y$.

\item If $Q=[\vx:\vA][y:I\vx]K$ with $K=(\vy:\vU)\st$ then
$\ps{Elim(I,Q,\va,c)\{\vf\}}= \se\ps\va\ps{c}\ps\vf$ where $\se$ is a
symbol of type $(\vx:\ps\vA)(y:\ps{I}\vx)(\vf:\ps\vT)\ps{K}$.

\item The translation of the other terms is defined recursively:
$\ps{uv}=\ps{u}\ps{v}$, \ldots
\end{lst}

\noindent
Let $\Up$ be the CAC whose symbols are $Ind_I, Constr^I_i, \we$ and
$\se$, and whose rules are:

\begin{rewc}
\we~Q~\vx~(Constr^I_i~\vz)~\vf & \D'_W[I,X,C_i,f_i,Q,\vf,\vz]\\
\se~\vx~(Constr^I_i~\vz)~\vf & \D'_S[I,X,C_i,f_i,Q,\vf,\vz]\\
\end{rewc}

\noindent
where $\D'_W[I,X,C,f,Q,\vf,\vz]$ and $\D'_S[I,X,C,f,Q,\vf,\vz]$ are
defined as follows:

\begin{lst}{--}
\item $\D'_W[I,X,X\vm,f,Q,\vf,\vz]= \D'_S[I,X,X\vm,f,Q,\vf,\vz]= f$,
\item $\D'_S[I,X,(z:B)D,f,Q,\vf,z\vz]=
\D'_S[I,X,D,f\,z,Q,\vf,\vz]$ and\\ $\D'_W[I,X,(z:B)D,f,Q,\vf,z\vz]=
\D'_W[I,X,D,f\,z,Q,\vf,\vz]$ if $X\notin\FV(B)$
\item $\D'_S[I,X,(z:B)D,f,Q,\vf,z\vz]=
\D'_S[I,X,D,f\,z\,[\vy:\vD]\se\vf\vq(z\vy),Q,\vf,\vz]$ and\\
$\D'_W[I,X,(z:B)D,f,Q,\vf,z\vz]=
\D'_W[I,X,D,f\,z\,[\vy:\vD]\we Q\vf\vq(z\vy),Q,\vf,\vz]$\\ if
$B=(\vy:\vD)X\vq$
\end{lst}

\noindent
Let $\thu$ be the typing relation of $\Up$.
\end{definition}


\begin{theorem}
The relation $\abip$ in CIC$^-$ preserves typing and is strongly
normalizing.
\end{theorem}

\begin{proof}
First, one can easily check that the translation preserves typing and
reductions:

\begin{lst}{--}
\item If $\G\th t:T$ then $\ps{\G}\thu \ps{t}:\ps{T}$.
\item If $\G\th t:T$ and $t\abip t'$ then $\ps{t}\a\ps{t'}$.
\end{lst}

\noindent
Thus, we are left to prove that $\Up$ satisfies the conditions of
Theorem \ref{thm-admis}. The symbols $\we$ and $\se$ are the canonical
recursors of $Ind_I$ w.r.t. the constructors $Constr^I_i$ (see
Definition \ref{def-can-rec}). Hence, subject reduction follows from
Lemma \ref{lem-sr}, and the fact that $\rec(Ind_I)=\{\we,\se\}$ is
admissible follows from Lemma \ref{lem-comp-acc} and Lemma
\ref{lem-head}.
\end{proof}


\section{Non-strictly positive types}
\label{sec-pos}

We are going to see that the use of elimination-based interpretations
allows us to have functions defined by recursion on non-strictly
positive types, while CIC has always been restricted to strictly
positive types. An interesting example is given by Abel's
formalization of first-order terms with continuations as an inductive
type $trm:\st$ with the constructors \cite{abel02tr}:

\begin{typc}
var & nat\A trm\\
fun & nat\A (list~trm)\A trm\\
mu & \non\non trm\A trm\\
\end{typc}

\noindent
where $list:\st\A\st$ is the type of polymorphic lists, $\neg X$ is an
abbreviation for $X\A\bot$ (in the next section, we will prove that
$\neg$ can be defined as a function), and $\bot:\st$ is the empty
type. Its recursor $rec:(A:\st)(y_1:nat\A A)$ $(y_2:nat\A list~trm\A
listA\A A)(y_3:\neg\neg trm\A \neg\neg A\A A)(z:trm)A$ can be defined
by the rules:

\begin{rewc}
rec~A~y_1~y_2~y_3~(var~n) & y_1~n\\
rec~A~y_1~y_2~y_3~(fun~n~l) & y_2~n~l~(map~trm~A~(rec~A~y_1~y_2~y_3)~l)\\
rec~A~y_1~y_2~y_3~(mu~f) &
y_3~f~[x:\neg A](f~[y:trm](x~(rec~A~y_1~y_2~y_3~y)))\\
\end{rewc}

\noindent
where $map: (A:\st)(B:\st)(A\A B)\A list~A\A list~B$ is defined by the
rules:

\begin{rewc}
map~A~B~f~(nil~A') & (nil~B)\\
map~A~B~f~(cons~A'~x~l) & cons~B~(f~x)~(map~A~B~f~l)\\
map~A~B~f~(app~A'~l~l') & app~B~(map~A~B~f~l)~(map~A~B~f~l')\\
\end{rewc}


We now check that $rec$ is an admissible recursor. Completeness
w.r.t. accessibility is easy. For the head-computability, we only
detail the case of $mu$. Let $f\s$, $t=mu~f\s$, $A\xi$, $A\t$ and
$\vy\t$ such that $\vide,\s\models\G_{mu}$ and $\xi,\s\t_z^t\models
\G=A:\st$, $\vy:\vU$ where $U_i$ is the type of $y_i$. Let
$b=recA\t\vy\t$, $c=[y:trm](x(by))$ and $a=[x:\neg A\t](f\s c)$. We
must prove that $y_3\t f\s a\in \I{A}_{\xi,\s\t_z^t}=A\xi$.

Since $\xi,\s\t_z^t\models \G$, $y_3\t\in \I{\neg\neg trm\A \neg\neg
A\A A}_{\xi,\t}$. Since $\vide,\s\models\G_{mu}$, $f\s\in \I{\neg\neg
trm}$. Thus, we are left to prove that $a\in \I{\neg\neg A}_{\xi,\t}$,
that is, $f\s c\g\in I_\bot$ for all $x\g\in \I{\neg
A}_{\xi,\t}$. Since $f\s\in \I{\neg\neg trm}$, it suffices to prove
that $c\g\in \I{\neg trm}$, that is, $x\g(by\g)\in I_\bot$ for all
$y\g\in I_{trm}$. This follows from the facts that $x\g\in \I{\neg
A}_{\xi,\t}$ and $by\g\in A\xi$ since $y\g\in I_{trm}$.

A general proof could certainly be given by using a general
formalization of inductive types like in \cite{matthes98thesis} for
instance.


\section{Inductive-recursive types}
\label{sec-indrec}

In this section, we define new positivity conditions for dealing with
{\em inductive-recursive type definitions} \cite{dybjer00jsl}. An
inductive-recursive type $C$ has constructors whose arguments have a
type $Ft$ with $F$ defined by recursion on $t:C$, that is, a predicate
$F$ and its domain $C$ are defined at the same time.

A simple example is the type $dlist:(A:\st)(\#:A\A A\A\st)\st$ of
lists made of distinct elements thanks to the predicate $fresh:
(A:\st)(\#:A\A A\A\st) A\A (dlist\,A\,\#)\A\st$ parametrized by a
function $\#$ to test whether two elements are distinct. The
constructors of $dlist$ are:

\begin{typc}
nil & (A:\st)(\#\!:\!A\!\A\! A\!\A\!\st)(dlist\,A\,\#)\\
cons & (A:\st)(\#\!:\!A\!\A\! A\!\A\!\st)
(x:A)(l:dlist\,A\,\#)(fresh~A~\#~x~l)\A (dlist\,A\,\#)\\
\end{typc}

\noindent
and the rules defining $fresh$ are:

\begin{rewc}
fresh~A~\#~x~(nil~A') & \top\\
fresh~A~\#~x~(cons~A'~y~l~h) & x\#y \et fresh~A~\#~x~l\\
\end{rewc}

\noindent
where $\top$ is the proposition always true and $\et$ the connector
``and''. Other examples are given by Martin-L\"of's definition of the
first universe {\em \`a la} Tarski \cite{dybjer00jsl} or by Pollack's
formalization of record types with manifest fields \cite{pollack02fac}.

For allowing defined predicate symbols in constructor types, we must
extend the notion of positive and negative positions by taking into
account the arguments in which a defined predicate symbol is monotone
or anti-monotone. We must also make sure that defined predicate
symbols are indeed monotone and anti-monotone in the arguments
declared to have this property.


\begin{definition}[Positive/negative positions - New definition]
\label{def-pos2}
Assume that every predicate symbol $f:(\vx:\vT)U$ with $U$ not a
product is equipped with a set $\mon^+(f)\sle A_f^\B=\{i\le |\vx|~|~
x_i\in\XB\}$ of {\em monotone arguments} and a set $\mon^-(f)\sle
A_f^\B$ of {\em anti-monotone arguments}. Definition \ref{def-pos} is
modified as follows:

\begin{lst}{--}
\item $\pos^\d(f\vt)= \{1^{|\vt|}~|~\d=+\}\cup
  \,\bigcup\{1^{|\vt|-i}2.\pos^{\ep\d}(t_i)~|~ \ep\in\{-,+\},\,
  i\in\mon^\ep(f)\}$,
\item $\pos^\d(tu)= 1.\pos^\d(t)$ if $t$ is not of the form $f\vt$.
\end{lst}
\end{definition}

For instance, in the positive type $trm$ of Section \ref{sec-pos},
instead of considering $\neg A$ as an abbreviation, one can consider
$\neg$ as a predicate symbol defined by the rule $\neg A\a A\A\bot$
with $\mon^-(\neg)=\{1\}$. Then, one easily check that $A$ occurs
negatively in $A\A\bot$, and hence that $trm$ occurs positively in
$\neg\neg trm$ since $\pos^+(\neg\neg trm)= \{1\}\cup 2.\pos^-(\neg
trm)= \{1\}\cup 2.2.\pos^+(trm)= \{1,2.2\}$.


\begin{definition}[Positivity conditions - New definition]
Definition \ref{def-pos-cond} is modified as follows. A pre-recursor
$f:(\vz:\vV)(z:C\vz)W$ is a {\em recursor} if:

\begin{lst}{--}
\item every $F\simeq C$ occurs only positively in $W$,
\item if $i\in\mon^\d(C)$ then $\pos(z_i,W)\sle \pos^\d(W)$.
\end{lst}

\noindent
Moreover, we assume that, for every rule $F\vl\a r\in\cR$ with
$F\in\cF^\B$:

\begin{lst}{--}
\item for all $i\in\mon^\ep(F)$, $l_i\in\XB$ and $\pos(l_i,r)\sle
  \pos^\ep(r)$.
\end{lst}
\end{definition}
 

Now, we must reflect these monotony properties in the
interpretations. Then, Theorem \ref{thm-admis} is still valid if we
prove that the interpretations for constant and defined predicate
symbols have all the monotony properties.

\begin{definition}[Monotone interpretation]
Let $\vS\le_i\vS'$ iff $S_i\le S_i'$ and, for all $j\neq i$,
$S_j=S_j'$. Let $F$ be a predicate symbol. An interpretation
$I\in\cR_\tF$ is {\em monotone} (resp. {\em anti-monotone}) {\em in
its $i$-th argument} if $I(\vt,\vS)\le I(\vt,\vS')$ whenever
$\vS\le_i\vS'$ (resp. $\vS\ge_i\vS'$). An interpretation $I\in\cR_\tF$
is {\em monotone} if it is monotone in every $i\in\mon^+(F)$ and
anti-monotone in every $i\in\mon^-(F)$. Let $\cR_\tF^m$ be the set of
monotone interpretations of $\cR_\tF$.
\end{definition}

One can easily check that $\cR_\tF^m$ is a complete lattice too. For
proving that interpretations for predicate symbols are monotone, we
need to prove Lemma \ref{lem-I-mon} again, and to prove a similar
lemma on candidate assignments.


\begin{lemma}
\label{lem-I-mon2}
If $I\le_f I'$, $\pos(f,t)\sle\pos^\d(t)$, $\G\th t:T$ and
$\xi\models\G$ then $\I{t}^I_{\xi,\t}\le^\d \I{t}^{I'}_{\xi,\t}$.
\end{lemma}

\begin{proof}
We only have to check the case $t=g\vt$. Let $R=\I{g\vt}^I_{\xi,\t}$
and $R'=\I{g\vt}^{I'}_{\xi,\t}$. $R=I_g(\vt\t,\vS)$ with
$\vS=\I{\vt}^I_{\xi,\t}$. $R'=I_g(\vt\t,\vS')$ with
$\vS'=\I{\vt}^{I'}_{\xi,\t}$. Let $i\le n=|\vt|$. If
$\pos(f,t_i)=\vide$ then $S_i=S_i'$. Otherwise, there is $\ep_i$ such
that $i\in\mon^{\ep_i}(f)$ and
$\pos(f,t_i)\sle\pos^{\ep_i\d}(t_i)$. Thus, by induction hypothesis,
$S_i\le^{\ep_i\d} S_i'$. Let $S_i^j=S_i$ if $i>j$, and $S_i^j=S_i'$
otherwise. $\vS^0=\vS$, $\vS^n=\vS'$ and, for all $j\le n$,
$\vS^{j-1}\le^{\ep_j\d}_j \vS^j$. Since $I_g$ is monotone, for all
$j\le n$, $I_g(\vt\t,\vS^{j-1})\le^{\ep_j^2\d} I_g(\vt\t,\vS^j)$, that
is, $I_g(\vt\t,\vS^{j-1})\le^\d I_g(\vt\t,\vS^j)$ since
$\ep_j^2=+$. Thus, $R=I_g(\vS)\le^\d I_g(\vS')$. Now, if $g\neq f$
then $I_g=I'_g$ and $R\le^\d R'$. If $g=f$ then $\d=+$ and $R\le R'$
since $I_f\le I'_f$.
\end{proof}


\begin{lemma}
\label{lem-xi-mon}
Let $\xi\le_x\xi'$ iff $x\xi\le x\xi'$ and, for all $y\neq x$,
$y\xi=y\xi'$. If $I$ is monotone, $\xi\le_x\xi'$, $x\in\pos^\d(t)$,
$\G\th t:T$ and $\xi,\xi'\models\G$ then $\I{t}^I_{\xi,\t}\le^\d
\I{t}^I_{\xi',\t}$.
\end{lemma}

\begin{proof}
By induction on $t$. The proof is very similar to the previous
lemma. We only detail the following two cases:

\begin{lst}{\bu}
\item $\I{x}^I_{\xi,\t}= x\xi\le x\xi'= \I{x}^I_{\xi,\t}$ and $\d=+$
necessarily.

\item Let $R=\I{g\vt}^I_{\xi,\t}$ and $R'=\I{g\vt}^I_{\xi',\t}$.
$R=I_g(\vt\t,\vS)$ with $\vS=\I{\vt}^I_{\xi,\t}$. $R'=I_g(\vt\t,\vS')$
with $\vS'=\I{\vt}^I_{\xi',\t}$. Let $i\le n=|\vt|$. If
$\pos(f,t_i)=\vide$ then $S_i=S_i'$. Otherwise, there is $\ep_i$ such
that $i\in\mon^{\ep_i}(f)$ and
$\pos(f,t_i)\sle\pos^{\ep_i\d}(t_i)$. Thus, by induction hypothesis,
$S_i\le^{\ep_i\d} S_i'$. Let $S_i^j=S_i$ if $i>j$, and $S_i^j=S_i'$
otherwise. $\vS^0=\vS$, $\vS^n=\vS'$ and, for all $j\le n$,
$\vS^{j-1}\le^{\ep_j\d}_j \vS^j$. Since $I_g$ is monotone, for all
$j\le n$, $I_g(\vt\t,\vS^{j-1})\le^{\ep_j^2\d} I_g(\vt\t,\vS^j)$, that
is, $I_g(\vt\t,\vS^{j-1})\le^\d I_g(\vt\t,\vS^j)$ since
$\ep_j^2=+$. Thus, $R\le^\d R'$.
\end{lst}
\end{proof}


\begin{lemma}
The interpretations for predicate symbols are monotone.
\end{lemma}

\begin{proof}
We first prove it for constant predicate symbols. Assuming that $I$ is
monotone, we must prove that $\vphi_C^I$ is monotone. Let
$i\in\mon^\d(C)$ and $\vS\le_i^\d\vS'$. We must prove that
$R=\vphi_C^I(\vt,\vS)\sle R'=\vphi_C^I(\vt,\vS')$. If some $t_i$ has
no normal form then $R=R'=\SN$. Assume now that every $t_i$ has a
normal form $t_i^*$. Let $t\in R$, $f\in\rec(C)$ of type
$(\vz:\vV)(z:C\vz)(\vy:\vU)V$, $\vy\xi$ and $\vy\t$ such that
$\xi_\vz^{\vS'},\t_\vz^\vt{}_z^t\models_I \vy:\vU$. We must prove that
$f\vt^* t\vy\t\in \I{V}_{\xi_\vz^{\vS'},\t_\vz^\vt{}_z^t}^I$. To this
end, it is sufficient to prove that
$\I{\vU}_{\xi_\vz^{\vS'},\t_\vz^\vt{}_z^t}^I\sle
\I{\vU}_{\xi_\vz^\vS,\t_\vz^\vt{}_z^t}^I$ and that
$\I{V}_{\xi_\vz^\vS,\t_\vz^\vt{}_z^t}^I\sle
\I{V}_{\xi_\vz^{\vS'},\t_\vz^\vt{}_z^t}^I$, which is the case by Lemma
\ref{lem-xi-mon} since $\pos(z_i,W)\sle\pos^+(W)$ by assumption.

We now prove that the interpretation for defined predicate symbols is
monotone. Let $F$ be a defined predicate symbol. Let $i\in\mon^\d(F)$
and $\vS\le_i^\d\vS'$. We must prove that $R=I_F(\vt,\vS)\sle
R'=I_F(\vt,\vS')$. Assume that every $t_i$ has a normal form $t_i^*$
and that $\vt^*=\vl\s$ for some rule $F\vl\a r\in\cR$. If this is not
the case then $R=R'=\SN$. So, $R=\I{r}^I_{\xi,\s}$ with
$x\xi=S_{\ka_x}$, and $R'=\I{r}^I_{\xi',\s}$ with
$x\xi'=S'_{\ka_x}$. If, for all $x\in\FVB(r)$, $\ka_x\neq i$, then
$\xi=\xi'$ and $R=R'$. Otherwise, $i=\ka_x$ for some $x$, and
$\xi\le^\d_x\xi'$. By Lemma \ref{lem-xi-mon}, $R\sle^{\d^2} R'$ since
$\pos(x,r)\sle\pos^\d(r)$ by assumption. Thus, $R\sle R'$ since
$\d^2=+$.
\end{proof}


\section{Conclusion}

By using an elimination-based interpretation for some inductive types,
we proved that the Calculus of Algebraic Constructions subsumes the
Calculus of Inductive Constructions almost completely. We define
general conditions on recursors for preserving strong normalization
and show that these conditions are satisfied by a large class of
recursors for strictly positive types and by some non-strictly
positive types too. Finally, we give general positivity conditions for
dealing with inductive-recursive types.\\

\noindent{\bf Acknowledgments.} I would like to thank very much
Christine Paulin, Ralph Matthes, Jean-Pierre Jouannaud, Daria
Walukiewicz-Chrz\k{a}szcz, Gilles Dowek and the anonymous referees for
their useful comments on previous versions of this paper. Part of this
work was performed during my stay at Cambridge (UK) in 2002 with Larry
Paulson thanks to a grant from the INRIA.


\end{document}